\documentclass[11pt,a4paper]{amsart}
\usepackage[svgnames,dvipsnames]{xcolor}
\usepackage{amssymb,amsmath,amsthm,enumerate}
\usepackage{hyperref}
\usepackage[left=2.5cm,right=2.5cm,top=2.5cm,bottom=3cm]{geometry}
\usepackage{caption}
\usepackage[scr=rsfso]{mathalfa}
\usepackage{upgreek}
\usepackage{nicefrac}
\usepackage{graphicx}
\usepackage{float}
\newcommand\myshade{90}
\colorlet{mylinkcolor}{violet}
\colorlet{mycitecolor}{YellowOrange}
\colorlet{myurlcolor}{Aquamarine}

\hypersetup{
  linkcolor  = mylinkcolor!\myshade!black,
  citecolor  = mycitecolor!\myshade!black,
  urlcolor   = myurlcolor!\myshade!black,
  colorlinks = true,
}

\usepackage{dsfont}

\newcommand{\pa}{\partial}

\newcommand{\eps}{\varepsilon}

\newcommand{\bR}{\mathbb{R}^3}

\newcommand{\bRfp}{\mathbb{R}^3 \times [0,\infty)}

\newcommand{\bS}{\mathbb{S}^2}

\newcommand{\mS}{\mathcal{S}}

\newcommand{\md}{\mathrm{d}}

\newcommand{\dpo}{d_{ij}(r,R)}
\newcommand{\dub}{d_{ij}(r,R) \, \tilde{b}_{ij}^{ub}(r, R) }

\newcommand{\g}{\gamma_{ij}}
\newcommand{\gi}{\bar{\bar{\gamma}}_{i}}

\newcommand{\gw}{\bar{\gamma}}
\newcommand{\gm}{\bar{\bar{\gamma}}}

\newcommand{\mP}{\mathrm{m}}

\newcommand{\la}{\langle}
\newcommand{\ra}{\rangle}

\newcommand{\sbar}{\bar{s}_{ij}}

\newcommand{\q}{q}
\newcommand{\qq}{q}
\newcommand{\pp}{p}

\newcommand{\Ls}{L^1}
\newcommand{\Linf}{L^\infty}

\newcommand{\cS}{\rho_{ij}}
\newcommand{\cLp}{c_{ij}}

\newcommand{\cgj}{c^j[g]}
\newcommand{\cfi}{c^i[f]}
\newcommand{\Cijfg}{  {C}_2^{ij}[f,g]}
\newcommand{\Cjigf}{  {C}^{ji}_2[g,f]}

\newcommand{\chfg}{{C}^{ij}_1[f,g]}
\newcommand{\chgf}{{C}^{ji}_1[g,f]}
\newcommand{\Bij}{  B_{ij}[f,g]}
\newcommand{\cchi}{\hat{c}_i}
\newcommand{\cchj}{\hat{c}_j}

\newcommand{\F}{\mathbb{F}}
\newcommand{\Q}{\mathbb{Q}}

\DeclareMathOperator{\esssup}{ess \, sup}

\allowdisplaybreaks

\newtheorem{theorem}{Theorem}

\newtheorem{lemma}[theorem]{Lemma}
\newtheorem{proposition}[theorem]{Proposition}

\newtheorem{remark}{Remark}

\begin{document}
	%%%%%%%%%%%%%%%%%%%%%%%%%%%%%%%%%%%%%%%%%%%%%%%%%%%%%%%%%%%%%%%%%%%%%

\bibliographystyle{abbrv}

	\title[Integrability propagation for  polyatomic gases]{ Integrability propagation for a Boltzmann system describing polyatomic gas mixtures}
	
		\author[R. Alonso]{Ricardo Alonso}
	\address{Texas A\&M, Division of arts \& sciences, Education City, Doha, Qatar }
	\email{ricardo.alonso@qatar.tamu.edu}

	\author[M. Pavi\'c-\v Coli\'{c}]{Milana Pavi\'c-\v Coli\'{c}}
	\address{Department of Mathematics and Informatics, 
		Faculty of Sciences, University of Novi Sad, 
		Trg Dositeja Obradovi\'ca 4, 21000 Novi Sad, Serbia}
	\address{
	Applied and Computational Mathematics,  RWTH Aachen University, Schinkelstr. 2, 52062
	Aachen, Germany }
	\email{milana.pavic@dmi.uns.ac.rs}

\begin{abstract}
This paper explores the $L^{p}$ Lebesgue's integrability propagation, $p\in(1,\infty]$, of a system of space homogeneous Boltzmann equations modelling a multi-component mixture of polyatomic gases based on the continuous internal energy.  For typical collision kernels proposed in the literature, $L^p$ moment-entropy-based estimates for the collision operator gain part and a lower bound for the loss part are performed leading to a vector valued inequality for the collision operator and, consequently, to a differential inequality for the vector valued solutions of the system. This allows to prove the propagation property of the polynomially weighted $L^p$ norms associated to the vector valued solution of the system of Boltzmann equations.  The case $p=\infty$ is found as a limit of the case $p<\infty$.
\end{abstract}

\maketitle

\medskip


	
%\tableofcontents	
	
\section{Introduction}
	
This paper is concerned with the Lebesgue's integrability propagation property of solutions for the system of Boltzmann-like equations describing a mixture of $P$ polyatomic gases.  The polyatomic model used in this document is based on the so-called continuous  approach established in \cite{DesMonSalv, LD-Toulouse, LD-Bourgat}, which introduces one additional (continuous) variable interpreted as an internal energy of a molecule that captures all peculiarities related to a more complex structure of a polyatomic molecule, like the non-translational  degrees of freedom in a collision.  Such a model is shown to be included in a kinetic modelling  general framework \cite{Bisi-Borsoni-Groppi} and to be accessible for a rigorous  analysis. For instance, the existence and uniqueness theory for the system of full non-linear Boltzmann equations  in the space homogeneous case is provided  in \cite{MPC-A-G}. For a single polyatomic gas, which is a subcase of the system considered in this paper, the global well-posedness for bounded mild solutions near global equilibria on torus is established in \cite{Duan-Li-poly}, while Fredholm property of a linearized  Boltzmann operator is studied in \cite{Borsoni-Comp, Brull-Comp, Brull-Comp-2, Bernhoff-lin-poly}  and recently in \cite{nicl-3} for  systems. \\

The system treated here is relevant for applications due to an increased need for accurate description of non-equilibrium processes in gas flows involving polyatomic molecules, mixtures and possible chemical reactions \cite{Kusto-book, Gio, Rugg-book-2}. At the formal level, improved continuum models that involve polyatomic gases can be obtained via various hydrodynamic asymptotics  of the Boltzmann equation, performed e.g. in  \cite{DesMonSalv,LD-ch-ens,Rugg-Bisi-6, Rugg-Arima-Sug,Bisi-Spiga-hydro,Groppi-Spiga,Anw-Bisi-Salv-Soa,SS-Anwasia,Bisi-Martalo,MPC}. The core issue is modelling of the  collision kernel of the Boltzmann polyatomic operator. A relevant test for the model is to compute transport coefficients and compare with experiments. Such a study is performed in \cite{MPC-Dj-T, MPC-Dj-T-O} for the single-species variant of the  collision kernel used in this paper and shows a perfect agreement with various physical requirements. \\

The propagation of $L^p$ integrability for solutions in the classical scalar Boltzmann equation has been addressed in works such as \cite{A,G,MV,AG,A1}.  This theory is important since it is the base for the study of Sobolev regularity propagation, long-time asymptotic theory of solutions for the Boltzmann equation and, consequently, the base for a rigorous derivation of macroscopical models such as the ones in the aforementioned paragraph.  It is also central in obtaining error estimates and convergence rates for numerical schemes resolving the model (which are usually based in a robust $L^{2}$-theory), see for example \cite{AGT-S}.  The analysis relies on integrability properties (bounds) satisfied by the gain collision operator analog to those of convolution operators Young's inequalities \cite{MV,GPV,AC,ACG} and a lower bound for the loss collision operator satisfied for solutions of the Boltzmann equation.  Conservation laws of mass, momentum, and energy are central to prove such bounds.  We stress here that estimates involving higher algebraic weights (associated to polynomial tails) require \textit{a priori} bounds on statistical moments, that is a robust $L^{1}_{k}$ theory.   Additionally, fine regularisation properties of the gain collisional operator are used in the aforementioned works such as the ones proven in \cite{L,BD,W}.  These regularisation properties (bounds) are technically difficult to establish and not yet available for the polyatomic operators used in this document.  It was shown in \cite{AGT}, applied to the Maxwell molecules case, that the $L^{p}$ integrability analysis of the Boltzmann equation can be simplified if the natural \textit{a priori} bound on the entropy is used appropriately.  The spirit of this latter approach is successfully used here in the polyatomic context, for more general collisional kernels, giving a complete description of higher integrability propagation in the full range of Lebesgue's integrability $p\in (1,\infty]$ for polyatomic mixtures.\\

Besides to the additional technical complexity of polyatomic collision operators, there are two important natural differences between the elastic scalar Boltzmann equation and the elastic system counterpart (monoatomic or polyatomic).  First, the symmetrisation of the scattering kernel, a common technique used in the scalar case to simplify the analysis, can not be used in collision operators associated to different species.  This is the case as different species are clearly not interchangeable from the collision mechanism's perspective.  Second, energy is interchanged among the species which bring interesting difficulties in finding lower bounds for the loss collision operators.  Indeed, it is difficult to establish in advance if a particular species loses all its energy in a particular point in time.  In practical terms, these two issues naturally lead to consider the system as a whole and seek inequalities for the vectorial distribution solution rather than for individual species.\\ 

The paper is organized as follows. Boltzmann collision operators and the corresponding system of Boltzmann equations describing a polyatomic gas mixture, together with the     functional setting are introduced in Preliminaries Section \ref{Sec: Prelim}. Assumptions on the collision kernel are listed in Section \ref{Section: assump coll kernel}. The crucial lemmas leading to the estimate of the collision operator are proven in Section \ref{Section: prelim lemmas}. Estimates on the gain part of the collision operator are given in Section \ref{Sec: gain}, whereas estimates for the loss part are proven in Section \ref{Sec: loss}.  Finally, these estimates allow to conclude on the bound of the pairwise collision operator written in a bi-linear form in Section \ref{Sec: coll op estimates} leading to the propagation result for polynomially weighted  $L^p$ norms associated to a vector-valued solution of the Boltzmann system proven in Section \ref{Sec: Prop}. 

\section{Preliminaries:   collision operators, Boltzmann system and  functional setting}\label{Sec: Prelim}

\subsection{Boltzmann collision operators modelling molecular interactions in a polyatomic gas mixture}
The underlying concept of the collision operator and the Boltzmann equation are collisions between molecules that change the state of a gas.  To efficiently describe molecular collisions in a multi-component polyatomic gas mixture, here assumed binary, fix any two mixture components $\mathcal{A}_i$ and $\mathcal{A}_j$,   $i,j\in \left\{ 1, \dots,  P \right\}$. Let the polyatomic molecule belonging to the species $\mathcal{A}_i$ have mass  $m_i$ and velocity-internal energy $(v', I')$, whereas molecule of species  $\mathcal{A}_j$ have mass  $m_j$ and velocity-internal energy $(v'_*, I'_*)$. After their collision, in the absence of chemical reactions, the two molecules will still belong to the same species with unchanged masses,  while velocity-internal energy  pairs change to  $(v, I)$ and $(v_*, I_*)$, respectively. During collision, conservation laws of 
momentum and the total (kinetic + microscopic internal) energy are  assumed conserved, that is 
\begin{equation}\label{p-p coll CL}
	m_i v' + m_j v'_*  = m_i v + m_j v_*, \qquad \frac{m_i}{2} |v'|^2 + I' +  \frac{m_j}{2} |v'_*|^2 + I'_* = \frac{m_i}{2} |v|^2 + I +  \frac{m_j}{2} |v_*|^2 + I_*.
\end{equation}
These laws can be rewritten in the center-of-mass framework, defined by 
  vectors of the center of mass velocity  $V_{ij}$, relative velocity $u$ and  reduced mass $\mu_{ij}$,
\begin{equation}\label{V-u}
	V_{ij}= \frac{m_i v + m_j v_*}{m_i + m_j}, \qquad u=v-v_*, \qquad 	\mu_{ij} = \frac{m_i m_j}{m_i + m_j}.
\end{equation}
Namely, \eqref{p-p coll CL} reduces to 
\begin{equation}\label{p-p en}
	V'_{ij} = V_{ij}, \quad E'_{ij} = E_{ij} := \frac{\mu_{ij}}{2} |u|^2 + I +  I_*.
\end{equation}  
It is usual to parametrize these equations with the scattering direction $\sigma\in\bS$ and energy exchange variables $r, R \in [0,1]$ by following the so-called  Borgnakke-Larsen procedure \cite{Bor-Larsen, LD-Bourgat}, allowing to  express pre--collisional velocities and internal energies in the original particle framework. Namely, first split the kinetic and internal energy part of the total energy $E'_{ij}=E_{ij}$ with  the variable $R \in [0,1]$,
\begin{equation}\label{R decom}
	\frac{\mu_{ij}}{2}  \left|u'\right|^2 = R E_{ij}, \quad  I' + I'_* =  (1-R) E_{ij}.
\end{equation}
Then, the kinetic part gives particles' velocities $v'$ and $v'_*$  by parametrizng the first equation with  $\sigma \in \bS$, while the total internal part is split on particles' internal energies $I'$ and $I'_*$ using the variable $r\in [0,1]$,
\begin{equation}\label{p-p coll rules}
	\begin{alignedat}{2}
		v'  &= V_{ij} + \frac{m_j}{m_i + m_j} \sqrt{\frac{2 \, R  \, E_{ij}}{\mu_{ij}}} \sigma,  \qquad
		I' &&=r (1-R) E_{ij}, \\
		v'_{*} &= V_{ij} - \frac{m_i}{m_i + m_j} \sqrt{\frac{2 \,  R \,  E_{ij}}{\mu_{ij}}} \sigma,
		\qquad I'_* &&= (1-r)(1-R) E_{ij}. 
	\end{alignedat}
\end{equation}
Last identities \eqref{p-p coll rules} allow to study the pre-post transformation 
\begin{equation}\label{pre-post T}
\mathcal{T}: (v, v_*, I, I_*, r, R, \sigma) \mapsto (v', v'_*, I', I'_*, r', R', \sigma'),
\end{equation}
where additionally to \eqref{p-p coll rules}, the primed variables  $\sigma' \in \bS$, $r', R' \in [0,1]$ are defined as 
\begin{equation}\label{p-p primed param}
	\sigma'=\frac{u}{\left|u\right|}, \qquad R'=\frac{\mu_{ij} \left|u\right|^2}{2 E_{ij}}, \qquad	r'=\frac{I}{I+I_*} = \frac{I}{E_{ij}-\frac{\mu_{ij}}{2}\left| u \right|^2}.
\end{equation} 
Transformation $\mathcal{T}$   is an involution  and its Jacobian, computed in \cite{DesMonSalv}, 
\begin{equation}\label{p-p Jac}
	J_{\mathcal{T}} : = \left| \frac{\pa(v', v'_{*},  I', \sigma', r' R')}{\pa (v,v_*,I_*,\sigma, r, R)  } \right| = \frac{(1-R) \sqrt{R}}{(1-R') \sqrt{R'}}.
\end{equation}
 will deeply influence the structure of collision operator.\\

In this paper, the collision operator that models an influence of the species $\mathcal{A}_j$ on the species $\mathcal{A}_i$,  for  distributions functions $f:= f(t,v,I)\geq 0$ and $g:= g(t,v,I)\geq0$  describing species $\mathcal{A}_i$ and $\mathcal{A}_j$, respectively,  is defined by
\begin{multline}\label{p-p coll operator}
	Q_{ij}(f,g)(v,I) = \int_{\bRfp} \int_{\bS \times [0,1]^2} \left\{ f(v',I')g(v'_*,I'_*) \left(\frac{I }{I' }\right)^{\alpha_i } \left(\frac{ I_*}{ I'_*}\right)^{ \alpha_j}   - f(v,I)g(v_*, I_*) \right\} \\ \times \mathcal{B}_{ij}(v, v_*, I, I_*, \sigma, r,  R) \, \dpo  \, \md \sigma \, \md r \, \md R \, \md v_* \, \md I_*,
\end{multline}
for $v'$, $v'_*$, $I'$ and $I'_*$    as in \eqref{p-p coll rules}. 
The same form of the collision operator is recently used in the Cauchy theory for the space homogeneous problem in \cite{MPC-A-G} and in the spirit of \cite{MPC-Dj-S} is a non-weighted  variant  of the one proposed in \cite{DesMonSalv}. We explain all involved terms in \eqref{p-p coll operator}. Parameters   $\alpha_i, \alpha_j >-1$ are here assumed constant, which corresponds to the so-called polytropic regime or calorically perfect gases. Theirs values are  related to  specific heats of   polyatomic gases and  can be matched to experiments \cite{MPC-SS-non-poly, MPC-Dj-T-O}.  In general, the transition probability  $	\mathcal{B}_{ij}$ is assumed to satisfy  the microreversibility relations  
\begin{equation}\label{p-p Bij micro}
	\mathcal{B}_{ij} :=	\mathcal{B}_{ij}(v, v_*, I, I_*, \sigma, r, R) = 	\mathcal{B}_{ij}(v', v'_*, I', I'_*, \sigma', r', R') = \mathcal{B}_{ji}(v_*, v, I_*, I, -\sigma, 1-r, R),
\end{equation}
corresponding to the interchange of pre-postcollisional molecules, and the interchange of colliding  molecules.  However, for the practical usage, usually some additional assumptions need to be imposed. In this paper, they are listed in the Section \ref{Section: assump coll kernel} and are taken from the existence theory \cite{MPC-A-G}.
Finally, the function
\begin{equation}\label{fun r R}
	\dpo =	  r^{\alpha_i}(1-r)^{\alpha_j} \, (1-R)^{\alpha_i + \alpha_j+1} \, \sqrt{R},
\end{equation}
is related to the  weight factor in the collision operator,  ensuring the invariance of
\begin{equation*}
	\left(	r (1-R) I \right)^{\alpha_i}  \left(	(1-r)(1-R) I_* \right)^{\alpha_j}  
\end{equation*}
  with respect to the changes described in \eqref{p-p Bij micro}. Together with   Jacobian \eqref{p-p Jac}, this implies the invariance of the measure 
\begin{equation}\label{meas inv}
	I^{\alpha_i} I_*^{\alpha_j} \dpo \, \md \sigma \, \md r \, \md R \, \md v_* \, \md I_* \, \md v \, \md I,
\end{equation}
with respect to changes from \eqref{p-p Bij micro}.

Under certain assumptions of integrability, which will be stated later, it is possible to see the collision operator  \eqref{p-p coll operator} as the difference between the gain part $	Q^+_{ij}(f,g)(v,I)$ and the loss part $	Q^-_{ij}(f,g)(v,I)$, where the gain operator is given by the integral form
 \begin{multline}\label{coll gain operator}
	Q^+_{ij}(f,g)(v,I) = \int_{\bRfp} \int_{\bS \times [0,1]^2}   f(v',I')g(v'_*,I'_*) \left(\frac{I }{I' }\right)^{\alpha_i } \left(\frac{ I_*}{ I'_*}\right)^{ \alpha_j}     \\ \times \mathcal{B}_{ij}(v, v_*, I, I_*, \sigma, r,  R) \, \dpo  \, \md \sigma \, \md r \, \md R \, \md v_* \, \md I_*,
\end{multline}
while the loss operator is local in $f$, more precisely,
\begin{equation}\label{coll loss operator}
	Q^-_{ij}(f,g)(v,I) =  f(v,I) \	\nu_{ij}[g](v,I),
\end{equation}
where the collision frequency $\nu_{ij}[g](v,I)$ is given by
\begin{equation}\label{coll fr}
	\nu_{ij}[g](v,I) = \int_{\bRfp} \int_{\bS \times [0,1]^2} g(v_*, I_*) \,  \mathcal{B}_{ij}(v, v_*, I, I_*, \sigma, r,  R) \, \dpo  \, \md \sigma \, \md r \, \md R \, \md v_* \, \md I_*.
\end{equation}

\subsubsection*{The collision operator weak form}
The invariance of the measure \eqref{meas inv} ensures   the well defined weak form of the collision operator \eqref{p-p coll operator}, namely
\begin{multline}\label{p-p weak form}
	\int_{\bRfp} Q_{ij}(f,g)(v, I) \, \chi(v,I) \, \md v\, \md I  
\\	=  \int_{(\bRfp)^2} \int_{\bS \times [0,1]^2} \left\{  \chi(v', I')  - \chi(v, I) \right\} \, f(v, I) \, g(v_*, I_*) 
	\\  \times \mathcal{B}_{ij}(v, v_*, I, I_*, r,  \sigma, R) \, \dpo \, \md \sigma \, \md r \, \md R \, \md v_* \, \md I_* \, \md v \, \md I,
\end{multline}
for any suitable test function  $\chi(v,I)$. Details can be found in \cite{DesMonSalv, MPC-A-G}.

\subsection{System of Boltzmann-like equations modelling a polyatomic gas mixture}

From the kinetic theory viewpoint, gas mixtures are modelled by associating a distribution function to each mixture component   and by assuming that all of them change due to mutual molecular interactions, here assumed to be binary collisions, captured with the collision operators. As these interactions can happen among molecules belonging to any  mixture  component, for the fixed species $\mathcal{A}_i$ the corresponding collision operators $Q_{ij}$ have to be summed over all possible partner components $\mathcal{A}_j$. Polyatomic nature of molecules is based on the continuous kinetic approach \cite{LD-Bourgat, LD-Toulouse, DesMonSalv}, meaning that each distribution function $f_i$, $i=1,\dots,P$, in the space homogeneous setting depends on time $t\geq 0$ and molecular velocity-internal energy pair $(v,I)$. Thus, the Boltzmann-like equation governing the evolution of each $f_i=f_i(t,v,I)$ reads
\begin{equation}\label{BE i}
	\partial_t f_i(t,v,I) = \sum_{j=1}^P Q_{ij}(f_i, f_j)(v,I), \quad i=1,\dots,P,
\end{equation}
where the collision operators are given in \eqref{p-p coll operator}. Gathering all such equations raises the system of Boltzmann equations that can be written in a vector form,
\begin{equation}\label{BE vector}
	\partial_t \F= \Q(\F,\F), \qquad 		\F =   \Big[f_i(t, v, I)\Big]_{i=1,\dots,P} , \ 	
	\Q(\F) =   \Big[ \sum_{j=1}^{P} Q_{ij}(f_i, f_j)(v, I) \Big]_{i=1,\dots,P}.
\end{equation}

\subsection{Functional spaces}

In this paper, we will work with the usual $L^p$ space 
\begin{equation}\label{Lp usual}
L^p = \left\{ \chi(v,I): \int_{\bRfp}  |\chi(v,I)|^p \md v\, \md I =:  \| \chi \|_{L^p}^p < \infty \right\},
\end{equation}
for $1\leq p < \infty$.
When it is important to highlight the space of integration we will emphasize the corresponding measure. For example, the norm of  $\chi$  with respect to $(v,I)$ will be denoted as
\begin{equation}\label{Lp (v,I)}
	\| \chi \|_{L^p(\md v \, \md I) } =  \left( \int_{\bRfp} |\chi|^p\, \md v\, \md I \right)^{1/p}.
\end{equation}

Polynomially weighted $L^p$ spaces will be of particular interest. First define  the Lebesgue brackets as in \cite{MPC-A-G},
	\begin{equation}\label{brackets}
	\la v, I \ra_i = \sqrt{1 + \frac{m_i}{2\, \mP}  |v|^2  + \frac{1}{ \mP}  I  },  \quad \text{for} \ i =1,\dots, P,
	\qquad \text{with} \quad	\mP =	\sum_{\ell=1}^{P} m_\ell.
\end{equation}
Weighted  $L^p$ spaces, with $1\leq p<\infty$, related to the species $i \in \left\{ 1, \dots, P  \right\}$ with  the weight associated to the Lebesgue brackets \eqref{brackets} of the order $k\geq 0$ is defined by
\begin{equation}\label{Lp i}
	L^p_{i, k} = \left\{ \chi(v,I): \int_{\bRfp} \left(  |\chi(v,I)| \la v, I \ra_i^{k} \right)^p \md v\, \md I = \| \chi \la \cdot \ra_i^{k} \|_{L^p}^p =:  \| \chi \|_{L^p_{i, k}}^p  < \infty \right\},
\end{equation}
while for $p=\infty$ the following definition is used
\begin{equation}\label{L inf i}
	L^\infty_{i, k} = \left\{ \chi(v,I): \esssup  |\chi(v,I)| \la v, I \ra_i^{k} =:  \| \chi \|_{L^\infty_{i, k}}  < \infty \right\}.
\end{equation}

Following \cite{Briant-Daus, Alonso-Orf} for the monatomic mixture case,   the vector valued distribution function that will model dynamics of  the whole mixture, is denoted by $\F(v,I)= \Big[f_i(v, I)\Big]_{i=1,\dots,P} $ . The related $L^p_k$  space is defined  for $ 1 \leq p  < \infty$ by
 \begin{equation}\label{Lp mixture}
	L^p_{k} =  \left\{ \F(v,I): 	\left\| \F \right\|_{L^p_{k}}^p :=   \sum_{i=1}^{P}  \int_{\bRfp}  \left(  |f_i(v, I)| \, \la v, I \ra_i^{k} \right)^p  \md v \, \md I  =  \sum_{i=1}^{P} \| f_i \|_{L^p_{i, k}}^p  < \infty \right\},
\end{equation}
while for $p=\infty$ is given by
 \begin{equation}\label{L infty mixture}
	L^\infty_{k} =  \left\{ \F(v,I): 	\left\| \F \right\|_{L^\infty_{k}}  :=  \sum_{i=1}^{P} \| f_i \|_{L^\infty_{i, k}}  < \infty \right\},
\end{equation}
for any $k \geq 0$.

 \section{Assumptions on the collision kernel}\label{Section: assump coll kernel}
 
 In this section, a rather general form of  the collision kernel is prescribed. It it worthwhile to highlight that  the collision kernel of this form is used in the existence theory \cite{MPC-A-G} and is compatible with recent studies of the linearized single-component Boltzmann operator \cite{Bernhoff-lin-poly,Brull-Comp-2,Duan-Li-poly}. Moreover, such a collision kernel is shown to be physically relevant in the case of single polyatomic gas endowed with frozen collisions \cite{MPC-Dj-T, MPC-Dj-T-O}.\\

 For any   $i, j \in \left\{ 1, \dots, P \right\}$, the transition probabilities  $\mathcal{B}_{ij}(v, v_*, I, I_*, \sigma, r,  R)\geq 0$ are assumed to satisfy the following bounds from above and below
 \begin{equation}\label{p-p ass B} 
 	b_{ij}(\hat{u}\cdot \sigma) \,  \tilde{b}_{ij}^{lb}(r, R) \,  \tilde{\mathcal{B}}_{ij}(v, v_*,I, I_*) \leq 	\mathcal{B}_{ij}(v, v_*, I, I_*, \sigma, r,  R) \leq   b_{ij}(\hat{u}\cdot \sigma) \,  \tilde{b}_{ij}^{ub}(r, R) \,  \tilde{\mathcal{B}}_{ij}(v, v_*,I, I_*),
 \end{equation}
 where the velocity-internal energy part $\tilde{\mathcal{B}}_{ij}(v, v_*,I, I_*)$ takes the following form
 \begin{equation}\label{p-p ass B tilde} 
 	\tilde{\mathcal{B}}_{ij}(v, v_*,I, I_*) % = \left( \frac{E_{ij}}{\mP} \right)^{\gamma_{ij}/2} 
 	= \left( \frac{1}{\mP} \left( \frac{\mu_{ij}}{2} |v-v_*|^2 + I + I_*  \right) \right)^{\gamma_{ij}/2} = \left( \frac{E_{ij}}{\mP} \right)^{\g/2}
 \end{equation}
 with  $E_{ij}$ defined in \eqref{p-p en} and 
 the rate $\gamma_{ij}$ satisfying
 \begin{equation}\label{gamma ij assumpt}
 	\gamma_{ij} = \gamma_{ji},\quad	\gamma_{ij} \in [0,2] \ \ \text{and, additionally,} \ \ \forall \ i \in \left\{ 1, \dots, P \right\} \quad  \max_{1 \leq j \leq P } \gamma_{ij} =: \gi >0.  
 \end{equation} 
 Note that this part of the collision kernel itself satisfies the micro-reversibility assumption \eqref{p-p Bij micro}.
 
 Non-negative  functions of energy exchange variables, $\tilde{b}_{ij}^{lb}(r, R)$ and $\tilde{b}_{ij}^{ub}(r, R)$, are assumed to have the following 
 %symmetry properties
 %\begin{equation*}
 %\tilde{b}_{ij}^{lb}(r, R) = \tilde{b}_{ji}^{lb}(1-r, R), \quad \tilde{b}_{ij}^{ub}(r, R) = \tilde{b}_{ji}^{ub}(1-r, R),
 %\end{equation*}
 %and 
 integrability properties
 \begin{equation}\label{bij r R integrable}
 	\tilde{b}_{ij}^{lb}(r, R), \ \tilde{b}_{ij}^{ub}(r, R)  \in L^1([0,1]^2; \,  \dpo \, \md r \, \md R),
 \end{equation}
 where the function $\dpo$ was introduced in \eqref{fun r R}.

 The angular part $b_{ij}(\hat{u}\cdot \sigma)\geq0$ is assumed non-negative and  symmetric with respect to the interchange $i\leftrightarrow j$, 
 \begin{equation}\label{bij integrable}
 	b_{ij}(\hat{u}\cdot \sigma) = b_{ji}(\hat{u}\cdot \sigma) \geq 0, \quad u=v-v_*, \quad \hat{u}=\frac{u}{|u|},
 \end{equation}
 and integrable on the unit sphere in $\bR$ denoted by $\bS$, that is, by means of spherical coordinates, 
 \begin{equation}\label{bij integrable 2}
 	\| b_{ij} \|_{L^1} = \int_{\bS} b_{ij}(\hat{u}\cdot\sigma) \, \md \sigma  = 2 \pi \int_{-1}^1 b_{ij}(x) \, \md x < \infty.
 \end{equation}

 For the future reference, we introduce here   decomposition of  the angular part of   collision kernel $b_{ij}(\hat{u}\cdot \sigma)$ depending on whether the argument is $\hat{u} \cdot \sigma \geq 0$ or $\hat{u} \cdot \sigma \leq 0$, namely
 \begin{equation}\label{bij decom}
 	b_{ij}(\hat{u}\cdot\sigma)  = b_{ij}(\hat{u}\cdot\sigma)  \, \mathds{1}_{\hat{u}\cdot \sigma \geq 0}  + b_{ij}(\hat{u}\cdot\sigma) \, \mathds{1}_{\hat{u}\cdot \sigma \leq 0} =:  b_{ij}^+(\hat{u}\cdot\sigma)   + b_{ij}^-(\hat{u}\cdot\sigma). 
 \end{equation}
 Due to the fact that in the mixture case collision operators are bilinear and symmetry properties of collision rules are more involved,  $b_{ij}(\hat{u}\cdot\sigma)$ cannot be symmetrized as it was done for the single monatomic gas Boltzmann equation without the loss of generality, and this decomposition is going to play an important role in the computations. The physical intuition is that  backward and forward scattering are not of the same nature for  components with different masses.

 \begin{remark}[On the possible choice of the collision kernel]
 	Since $E_{ij}$ defined in \eqref{p-p en} is micro-reversible in the sense of \eqref{p-p Bij micro}, one possible  choice for the collision kernel $\mathcal{B}_{ij}$ is
 	\begin{equation}\label{Bij choice energy}
 		\mathcal{B}_{ij}(v, v_*, I, I_*, \sigma, r,  R)  =    b_{ij}(\hat{u}\cdot \sigma) \,  \left( \frac{E_{ij}}{\mP} \right)^{\g/2},
 	\end{equation}
 	with $\g$ and $b_{ij}$ satisfying assumptions \eqref{gamma ij assumpt} and \eqref{bij integrable}-\eqref{bij integrable 2} above. 
 \end{remark}

 It is convenient to introduce here  notation that will be used in the rest of the paper. 
 For the rate $\g$, the following notation will be used
 \begin{equation}\label{gamma w}
 	\gw := \min_{1 \leq i \leq P } \gi   >0, \qquad \gm :=  \max_{1 \leq i \leq P } \gi   >0.
 \end{equation} 
 Moreover, for  mass ratios related to the fixed pair of particles $\mathcal{A}_i$ and $\mathcal{A}_j$,  the following parameter is introduced
 \begin{equation}\label{s min}
 	\sbar = \min \left\{ \frac{m_i}{m_i + m_j}, \frac{m_j}{m_i + m_j}  \right\}.
 \end{equation}

\subsection{Solutions to the Boltzmann system of equations}\label{Sec: solutions}

In this paper we will work with solutions to the Boltzmann system of equations \eqref{BE vector} in the space homogeneous setting with initial data
\begin{equation}\label{BE in}
	\F(0, v, I)=\F_0(v,I).
\end{equation}
The Cauchy problem \eqref{BE vector} and \eqref{BE in} is resolved in   \cite{MPC-A-G} in the context of the Boltzmann system describing a complex multi-component mixture composed of $M$ monatomic and $P$ polyatomic gases introduced in \cite{LD-ch-ens}. Note that the present  
 mixture of solely polyatomic gases fits into this framework, since it can be understood as  a subsystem obtained by  taking  the number of monatomic gases to be zero,   $M=0$. 
 
 More precisely, it is shown that \eqref{BE vector} and \eqref{BE in}  with the collision kernel satisfying assumptions \eqref{p-p ass B} has a unique solution in $\mathcal{C}([0,\infty), \Omega)  \cap \mathcal{C}^1((0,\infty), L^1_2)$, provided that the initial data $\F_0$ belong to the set $\Omega   \subseteq L^1_2$, that in our notation reads, 
\begin{equation}\label{Omega}
	\Omega = \left\{ \F(v,I) \in L^1_2: \ \F \geq 0, \ 0< \| f_i \|_{L^1_{i,0}}  < \infty, \
\| \F \|_{L^1_{(2+\gm - \gw)^+}}   < \infty \right\},
\end{equation}
with $\gw$ and $\gm $  from \eqref{gamma w}.

Moreover,   such a solution generates and propagates  polynomial moments, which are $L^1_k$ norms in our notation \eqref{Lp mixture}. As in this paper we study more general  $L^{p}_k$ Lebesgue's integrability propagation, it is particularly relevant to state the $L^1_k$ propagation result from \cite{MPC-A-G}.
 
\begin{proposition}[Propagation of  $L^1_k$  moments from \cite{MPC-A-G}]\label{Prop L1}
	Let $\F=[f_i]_{1\leq i \leq P} \geq 0$ be the solution of the system of Boltzmann equations \eqref{BE vector}  with the collision kernel    \eqref{p-p ass B}. Then, if  $	\| \F \|_{L^1_{k}}(0) < \infty$,   for any $k  > 2$,  the following estimate holds
	\begin{equation}\label{poly prop}
		\| \F \|_{L^1_{k}}(t) \leq \max \left\{ 	\mathcal{E}_k, e	\| \F \|_{L^1_{k}}(0) \right\}, \qquad \forall t\geq 0,
	\end{equation}
where the constant $\mathcal{E}_k$ is explicitly computed in \cite{MPC-A-G} and depends on $k$, the choice of the collision kernel and initial data.
 \end{proposition}

\section{Towards  the estimate on the collision gain operator}\label{Section: prelim lemmas}

This section establishes fundamental lemmas that will allow to estimate the gain part of the collision operator \eqref{coll gain operator}.

\subsection{Estimates on the averaging operators}
For the angular part of the collision kernel $b_{ij}$ decomposed as in \eqref{bij decom}, we define the following averaging operators over the space of collision variables -- scattering direction $\sigma \in \bS$ and energy exchange variables $r, R \in[0,1]$,
\begin{equation}\label{S operators}
	\begin{split}
		\mS^{\pm}_{ij}(\chi)(v, I, v_*, I_*) &= \int_{\bS \times [0,1]^2} \chi(v', I') \, b_{ij}^{\pm}(\hat{u}\cdot\sigma) \, r^{-\frac{\g}{2 \qq}} \,  \dub \, \md r \, \md R \, \md \sigma,
	\end{split}
\end{equation}
where    $v', I'$ are given in \eqref{p-p coll rules}, $\g \in [0,2]$, $d_{ij}$ is the measure defined by \eqref{fun r R}, $\tilde{b}_{ij}^{ub}$ comes from the assumption on the collision kernel \eqref{p-p ass B},  and $\qq\geq 1$ is such that the following constant $\cS$ is finite
\begin{equation}\label{cond sing r}
	\cS(\qq) = \int_{[0,1]^2} r^{-\left(1+\frac{\g}{2}\right)\frac{1}{\qq}}(1-R)^{-\frac{1}{\qq}} \,  \dub \, \md r \, \md R < \infty.
\end{equation}
In the case of constant $\tilde{b}_{ij}^{ub}$, the value of $\qq$   for which the theory holds is discussed in Remark \ref{Remark const S op}.

For the system analysis we highlight importance of the decomposition \eqref{bij decom} of angular part $b_{ij}$ related to the lack of interaction symmetry noticed in the monatomic mixture case \cite{Alonso-Orf} where is essential to differentiate the forward scattering $ \hat{u}\cdot \sigma \geq 0 $ from the backward scattering $\hat{u}\cdot \sigma \leq 0$ in $\sigma$, as the species are not interchangeable.  Moreover, the test function $\chi$ being evaluated at the post-collisional state induces the pre-post transformation that brings different bounds depending on these two cases. The same holds for the classical case of  monatomic Boltzmann operator. However, the difference between the two models lies in the appearance of a singularity in the involved Jacobian related to the internal energy: the polyatomic nature of the model reflected through the functional space with an additional integration over the internal energy variable removes the classical singularity in $\hat{u}\cdot \sigma$ appearing in monatomic model \cite{IG-Alonso-BAMS} or even monatomic mixture model \cite{Alonso-Orf}, at the price of having the new condition \eqref{cond sing r} related only to the energy exchange variables.\\

The upcoming lemmas \ref{Lemma S operators} and \ref{Lemma S operators bdd}, reminiscent of \cite[Proposition 4.2]{GPV} study $L^p$ norms of averaging operators \eqref{S operators}, when $b_{ij}$ belongs to $L^1$ and $L^\infty$, respectively. When the angular part is assumed integrable,  $L^p$ norms can be taken only over specific variables, while this rule will be weakened for bounded angular parts.

\begin{lemma}\label{Lemma S operators} 
	Assume $b_{ij}$ decomposed as in \eqref{bij decom} belongs to  $\Ls$. Then for $\qq\geq1$ such that the constant \eqref{cond sing r} is finite and  for a suitable test function $\chi \in L^{\qq}$, the following estimates hold on the operators defined by \eqref{S operators},
	\begin{description}
		\item[\textit{(a)}]  For the operator $	\mS^+_{ij} $,
		\begin{equation}\label{S+ estimate L1}
			\sup_{(v_*, I_*)}	\left\| 	\mS^+_{ij}(\chi) \right\|_{L^{\qq}(\md v\, \md I)}  \leq 2^{1/(2\qq)}  \left( \sbar \right)^{-3/\qq} \, \cS(\qq) \,   \| b_{ij}^+ \|_{\Ls}   \| \chi\|_{L^{\qq}},
		\end{equation}
		\item[\textit{(b)}]  For the operator $	\mS^-_{ij} $,
	\begin{equation}\label{S- estimate L1}
		\sup_{(v, I)}	\left\| 	\mS^-_{ij}(\chi) \right\|_{L^{\qq}(\md v_*\, \md I_*)}  \leq 2^{1/(2\qq)}  \left(  \sbar \right)^{-3/\qq} \, \cS(\qq) \,   \| b_{ij}^- \|_{\Ls}   \| \chi\|_{L^{\qq}},
	\end{equation}
	\end{description}
where $\sbar$ is defined in \eqref{s min}.
\end{lemma}

\begin{proof} Let us  first prove the part (a). 
We   apply  Minkowski's integral inequality  and then for the angular integration we use  H\"older's inequality, which  together with the fact that $b_{ij}^+$ is defined for $\hat{u}\cdot\sigma \geq 0$ and Fubini's theorem yields
\begin{multline}\label{S+ est 1}
\left\| 	\mS^+_{ij}(\chi) \right\|_{L^{\qq}(\md v\, \md I)}
\\
\leq \int_{[0,1]^2} \left(  \int_{\bRfp} \left|  \int_{\bS} \chi(v', I') \, b_{ij}^+(\hat{u}\cdot \sigma) \, \md \sigma \right|^{\qq} \md v \, \md I  \right)^{1/\qq} r^{-\frac{\g}{2 \qq}} \dub \, \md r \, \md R
 \\
\leq  \| b_{ij}^+ \|_{\Ls}^{1/\pp} \int_{[0,1]^2} \left(  \int_{\bS} \int_{\bRfp}  \left| \chi(v', I') \right|^{\qq} \, b_{ij}^+(\hat{u}\cdot \sigma) \,  \md v \, \md I   \, \md \sigma \right)^{1/\qq} r^{-\frac{\g}{2 \qq}} \dub \, \md r \, \md R.
\end{multline}
Fixing parameters $(r, R, \sigma)$,   consider the   change  of variables $(v, I) \mapsto (v', I')$ given by \eqref{p-p coll rules}. Its   Jacobian is similar to the one for a single polyatomic gas \cite{Brull-Comp-2},
\begin{equation}\label{Jac}
\left| \frac{ \partial(v', I')}{\partial(v, I)} \right|= \left( \frac{m_i}{m_i+m_j} \right)^3  r \, (1-R).
\end{equation}
We mention that the Jacobian \eqref{Jac} does not have singularity in $\hat{u}\cdot\sigma$, as it was the case for  a single component monatomic gas. As a matter of fact, the energy variable $I$ helps to remove the singularity, but, as we will see later, it brings another difficulty related to the  energy exchange variables. 

Now, 
the argument of   collision kernel angular part $\hat{u}\cdot \sigma \geq0$ needs to be expressed  in terms of new variables. To that aim,  recall  $v'$ from \eqref{p-p coll rules} and define
\begin{equation}\label{v-v_*}
w_* :=	v'-v_*  = \frac{m_i}{m_i+m_j} u   +  \frac{m_j}{m_i+m_j} \sqrt{\frac{2 \, R }{\mu_{ij}} \frac{I'}{r(1-R)} } \ \sigma,
\end{equation}
which allows to express 
\begin{equation}\label{u +}
\frac{m_i}{m_i+m_j} u =  w_*  -  a\, |w_*|  \, \sigma, \quad \text{with} \ a :=  \frac{m_j}{m_i+m_j} \sqrt{\frac{2 \, R }{\mu_{ij}}  \frac{I'}{r(1-R)} } \frac{1}{|w_*|},
\end{equation}
implying 
\begin{equation*}
	\frac{m_i}{m_i+m_j}\frac{ |u| }{|w_*|} = \left|  \hat{w}_*  - a\, \sigma \right| = \sqrt{1 + a^2 - 2 \, a \, \hat{w}_* \cdot \sigma }.
\end{equation*}
Thus,
scalar  multiplication of \eqref{v-v_*}    with $\sigma$ yields
\begin{equation}\label{v'-v_* +}
\hat{w}_* \cdot \sigma = \left( \sqrt{1 + a^2 - 2 \, a \, \hat{w}_* \cdot \sigma } \right)\hat{u} \cdot \sigma + a \quad \Rightarrow \quad 
\hat{u} \cdot \sigma = \frac{\hat{w}_*\cdot \sigma-a}{{\sqrt{1 + a^2 - 2 \, a \, \hat{w}_*\cdot \sigma} }}.
\end{equation}
Therefore, we can finalize the change of variables \eqref{Jac} and for the inner integral of \eqref{S+ est 1} write, after  Fubini's theorem,
\begin{multline}\label{pomocna 1}
 \int_{\bS \times \bRfp }  \left| \chi(v', I') \right|^{\qq} \, b_{ij}^+(\hat{u}\cdot \sigma)  \,  \md v \, \md I \, \md \sigma
%	 \\	  =\left( \frac{m_i}{m_i+m_j} \right)^{-3}  \frac{1}{r \, (1-R)}   \int_{\bRfp} \left| \chi(v', I') \right|^{\qq}   \int_{\bSp} b_{ij}^+(x)\, \md \sigma \,  \md v' \, \md I'   
	  \\
	    =\left( \frac{m_i}{m_i+m_j} \right)^{-3}  \frac{1}{r \, (1-R)}     \int_{\bRfp}  \left| \chi(v', I') \right|^{\qq} \,  \mathcal{I}_+  \, \md v' \, \md I'
	\\    \leq  \left( \sbar \right)^{-3}  \frac{1}{r \, (1-R)}     \int_{\bRfp}  \left| \chi(v', I') \right|^{\qq} \,  \mathcal{I}_+  \, \md v' \, \md I'\, 
\end{multline}	    
with $\sbar$ from \eqref{s min} and 
\begin{equation}
\mathcal{I}_+ = \int_{ \bS}  b_{ij}^+\Big( \tfrac{\hat{w}_*\cdot \sigma-a}{{\sqrt{1 + a^2 - 2 a \, \hat{w}_*\cdot \sigma} }} \Big)\, \md \sigma ,
\end{equation}
keeping in mind $w_*=v'-v_*$ as in \eqref{v-v_*}, $a$ given  in  \eqref{u +}, and   $b_{ij}^+$ is defined for a positive argument.
%and the half sphere defined now for $\hat{u}\cdot \sigma$ expressed as in \eqref{x y plus}. 
The next aim is to integrate with respect to $\sigma$ and, in fact, to estimate the integral $\mathcal{I}_+$ in terms of $L^1$-norm of $b_{ij}^+$. To that end, we use spherical coordinates with the direction of $\hat{w}_*$, and, as there is no other dependency on $\sigma$ in the expression \eqref{pomocna 1}, for the azimuthal angle $\theta$ we take the angle between direction $\hat{w}_*$ and $\sigma$, allowing to pass to the integration with respect to $y= \hat{w}_* \cdot \sigma = \cos \theta$,
\begin{equation*}
		\mathcal{I}_+ = 2 \pi \int_0^\pi b_{ij}^+\Big( \tfrac{\cos\theta-a}{{\sqrt{1 + a^2 - 2 a \, \cos\theta} }} \Big) \sin\theta \, \md \theta
			=  2 \pi \int_{-1}^1 b_{ij}^+\Big( \tfrac{y-a}{{\sqrt{1 + a^2 - 2 a y} }} \Big)   \md y.
	\end{equation*} 
Next, we change variables $y \mapsto x$ given by
\begin{equation}\label{x y plus}
	y=\sqrt{ 1 + a^2 -2\,a \,y } \  x+ a \quad \Rightarrow \quad	x= \frac{y-a}{{\sqrt{1 + a^2 - 2 a y} }},
\end{equation}
which is exactly \eqref{v'-v_* +} written in terms of $x=\hat{u}\cdot \sigma$ and $y=\hat{w}_*\cdot \sigma$. Since $b^+_{ij}$ is only supported in $ \Big\{ y: \tfrac{y-a}{{\sqrt{1 + a^2 - 2 a y} }}  \in [0,1]\Big\}$, it implies $x\in[0,1]$ and  $0\leq a \leq y \leq 1$. The Jacobian is computed by implicit differentiation of \eqref{x y plus} 
\begin{equation*}
	\frac{\md y}{\md x} = \frac{1 + a^2 - 2 a y}{\sqrt{1 + a^2 - 2 a y} + a x} \leq \sqrt{1 + a^2 - 2 a y} \leq \sqrt{1 + a^2 } \leq \sqrt{2},
\end{equation*}
and the estimate is due to the domain.
These considerations provide the estimate for  $\mathcal{I}_+ $, namely
\begin{equation*}
	\mathcal{I}_+ \leq \sqrt{2} \, 2 \pi \int_{x\in[0,1]} b_{ij}^+(x)  \md x =  \sqrt{2} \,  \| b^+_{ij} \|_{\Ls}.
\end{equation*} 
Note that this estimate is particularly elegant in three dimensional space.

Returning to the inner integral \eqref{pomocna 1}, this gives
\begin{equation*}
	 \int_{\bS}	 \int_{\bRfp}  \left| \chi(v', I') \right|^{\qq} \, b_{ij}^+(\hat{u}\cdot \sigma)  \,  \md v \, \md I \, \md \sigma  
%	 \\
%	  \leq \left( \frac{m_i}{m_i+m_j} \right)^{-3}  \frac{\sqrt{2}}{r \, (1-R)}   \| \chi\|_{L^{\qq}}^{\qq} \  \| b_{ij}^+ \|_{\Ls}
	    \\
	  \leq \left( \sbar \right)^{-3}  \frac{\sqrt{2}}{r \, (1-R)}   \| \chi\|_{L^{\qq}}^{\qq} \  \| b_{ij}^+ \|_{\Ls}.
\end{equation*}
Therefore, for any $(v_*, I_*)$, \eqref{S+ est 1}  is estimated as 
 \begin{equation}\label{S+ est 2}
 	\left\| 	\mS^+_{ij}(\chi) \right\|_{L^{\qq}(\md v\, \md I)}\leq 2^{1/(2\qq)}  \left( \sbar \right)^{-3/\qq} \, \cS(\qq) \,   \| b_{ij}^+ \|_{\Ls}   \| \chi\|_{L^{\qq}},
\end{equation}
implying \eqref{S+ estimate L1} and concluding the part (a). \\

The part (b) follows the same strategy. Firstly exploiting   Minkowski and  H\"older inequalities and    Fubini's theorem, the following estimate is obtained 
\begin{multline}\label{S- est 1}
	\left\| 	\mS^-_{ij}(\chi) \right\|_{L^{\qq}(\md v_*\, \md I_*)}
%	\\
%	\leq \int_{[0,1]^2} \left(  \int_{\bRfp} \left|  \int_{\bSp} \chi(v', I') \, b_{ij}^+(\hat{u}\cdot \sigma) \, \md \sigma \right|^{\qq} \md v \, \md I  \right)^{1/\qq} r^{-\frac{\g}{2 \qq}} \dub \, \md r \, \md R
	\\
	\leq  \| b_{ij}^- \|_{\Ls}^{1/\pp} \int_{[0,1]^2} \left(  \int_{\bS} \int_{\bRfp}  \left| \chi(v', I') \right|^{\qq} \, b_{ij}^-(\hat{u}\cdot \sigma) \,  \md v_* \, \md I_*   \, \md \sigma \right)^{1/\qq} r^{-\frac{\g}{2 \qq}} \dub \, \md r \, \md R.
\end{multline}
Now for the fixed parameters $(r, R, \sigma)$,  he   change  of variables $(v_*, I_*) \mapsto (v', I')$ is performed, with Jacobian
\begin{equation}\label{Jac star}
	\left| \frac{ \partial(v', I')}{\partial(v_*, I_*)} \right|= \left( \frac{m_j}{m_i+m_j} \right)^3  r \, (1-R).
\end{equation}
The argument $\hat{u}\cdot \sigma \leq 0$ is expressed in terms of new variables by recalling equation for $v'$ from \eqref{p-p coll rules} and denoting
\begin{equation}\label{v'-v}
w:=	v'-v   = \frac{m_j}{m_i+m_j}  \left( - u +  \sqrt{\frac{2 \, R }{\mu_{ij}} \frac{I'}{r(1-R)} }  \sigma \right).
\end{equation}
This yields expression for $u$, 
\begin{equation}\label{u -}
	\frac{m_j}{m_i+m_j} u =  - w  +  a\, |w|  \, \sigma, \quad \text{with} \ a :=  \frac{m_j}{m_i+m_j} \sqrt{\frac{2 \, R }{\mu_{ij}}  \frac{I'}{r(1-R)} } \frac{1}{|w|},
\end{equation}
and allows to express 
\begin{equation*}
	\frac{m_j}{m_i+m_j}\frac{ |u| }{|w|} = \left|  \hat{w}  - a\, \sigma \right| = \sqrt{1 + a^2 - 2 \, a \, \hat{w} \cdot \sigma }.
\end{equation*}
Thus, taking the scalar product  of \eqref{v'-v}   with $\sigma$ yields
\begin{equation}\label{v'-v -}
	\hat{w} \cdot \sigma = - \left( \sqrt{1 + a^2 - 2 \, a \, \hat{w}_* \cdot \sigma } \right)\hat{u} \cdot \sigma + a \quad \Rightarrow \quad 
	\hat{u} \cdot \sigma = - \frac{\hat{w}\cdot \sigma-a}{{\sqrt{1 + a^2 - 2 \, a \, \hat{w}\cdot \sigma} }}.
\end{equation}
Therefore, the inner integral in \eqref{S- est 1} can be rewritten using the change of variables \eqref{Jac star} as follows
\begin{multline}\label{pomocna 2}
	 \int_{\bS \times \bRfp}  \left| \chi(v', I') \right|^{\qq} \, b_{ij}^-(\hat{u}\cdot \sigma)  \,  \md v_* \, \md I_* \, \md \sigma
	%	 \\	  =\left( \frac{m_i}{m_i+m_j} \right)^{-3}  \frac{1}{r \, (1-R)}   \int_{\bRfp} \left| \chi(v', I') \right|^{\qq}   \int_{\bSp} b_{ij}^+(x)\, \md \sigma \,  \md v' \, \md I'   
	\\
	=\left( \frac{m_j}{m_i+m_j} \right)^{-3}  \frac{1}{r \, (1-R)}   \int_{\bRfp} \left| \chi(v', I') \right|^{\qq} 	\mathcal{I}_-  \ \md v' \, \md I'
	\\
	\leq \left(\sbar \right)^{-3}  \frac{1}{r \, (1-R)}   \int_{\bRfp} \left| \chi(v', I') \right|^{\qq} 	\mathcal{I}_-  \ \md v' \, \md I',
\end{multline}	
with  $\sbar$ from \eqref{s min} and
\begin{equation*}
	\mathcal{I}_- = \int_{ \bS} b_{ij}^-\Big( \tfrac{-\hat{w}\cdot \sigma + a}{{\sqrt{1 + a^2 - 2 \, a \, \hat{w}\cdot \sigma} }} \Big)\, \md \sigma.
\end{equation*}
Using the same ideas as in the previous case, the angular integral $	\mathcal{I}_-$ can be estimated using the spherical coordinates with $\hat{w}$ as direction, and  the angle $\theta$ between $\hat{w}$ and $\sigma$ as azimuth,   that turn the integration with respect to half sphere into the one dimensional integration with respect to $y=\cos\theta = \hat{w}\cdot\sigma$, more precisely,
\begin{equation*}
	\mathcal{I}_- =	 2\pi  \int_0^\pi b_{ij}^-\Big( \tfrac{-\cos\theta+a}{{\sqrt{1 + a^2 - 2 \, a \, \cos\theta} }} \Big) \sin\theta \md \, \theta 
	=  2\pi \int_{  -1}^1 b_{ij}^-\Big( \tfrac{-y+a}{{\sqrt{1 + a^2 - 2 \, a \, y} }} \Big) \md  y.
\end{equation*}
Finally, we change variables $y \mapsto x$, given by
\begin{equation}\label{x y -}
	y= - \sqrt{ 1 + a^2 -2\,a \,y } \  x+ a \quad \Rightarrow \quad		x= - \frac{y-a}{{\sqrt{1 + a^2 - 2 a y} }} \in[-1, 0].
\end{equation}
Since $0\leq a \leq y \leq 1$, the Jacobian of this transformation can be estimated as
\begin{equation*}
	\left| \frac{\md y}{\md x} \right| =   \frac{ 1 + a^2 - 2 a y }{\sqrt{1 + a^2 - 2 a y} - a x}   \leq \sqrt{1 + a^2 - 2 a y} \leq \sqrt{1 + a^2 } \leq \sqrt{2},
\end{equation*} 
leading to 
\begin{equation*}
	\mathcal{I}_- \leq \sqrt{2} \, 2 \pi \int_{x\in[-1,0]} b_{ij}^-(x)  \md x =  \sqrt{2} \,  \| b^-_{ij} \|_{\Ls}.
\end{equation*}                                                                                                                
%\begin{equation}\label{v'-v +}
%	w \cdot \sigma =  \frac{m_j}{m_i + m_j} \left( - u \cdot \sigma +       \sqrt{\frac{2 \, R }{\mu_{ij}} \frac{I'}{r(1-R)} } \right).
%\end{equation}
%For the case under consideration $\hat{u}\cdot\sigma \leq 0$, i.e. $\sigma \in \bSm$, we are led to introduce variables 
% \begin{equation*}
% 	x= \hat{u} \cdot \sigma \in [-1, 0], \quad y= \hat{w} \cdot \sigma \in [0,1], \quad \text{and} \quad  a =  \frac{m_j}{m_i+m_j} \sqrt{\frac{2 \, R }{\mu_{ij}}  \frac{I'}{r(1-R)} } \frac{1}{|w|} \in [0,1].
% \end{equation*}
 Thus, \eqref{pomocna 2} becomes
\begin{equation*}	
	\int_{\bS} \int_{\bRfp}  \left| \chi(v', I') \right|^{\qq} \, b_{ij}^-(\hat{u}\cdot \sigma)  \,  \md v_* \, \md I_* \, \md \sigma	
%	\leq \left( \frac{m_j}{m_j+m_j} \right)^{-3}  \frac{\sqrt{2}}{r \, (1-R)}   \| \chi\|_{L^{\qq}}^{\qq} \  \| b_{ij}^- \|_{\Ls}
%	\\
	\leq \left( \sbar \right)^{-3}  \frac{\sqrt{2}}{r \, (1-R)}   \| \chi\|_{L^{\qq}}^{\qq} \  \| b_{ij}^- \|_{\Ls}
\end{equation*}
which for \eqref{S- est 1} implies, for any $(v,I)$,
\begin{equation}
	\left\| 	\mS^-_{ij}(\chi) \right\|_{L^{\qq}(\md v_*\, \md I_*)}   \leq 2^{1/(2\qq)}  \left( \sbar \right)^{-3/\qq} \, \cS(\qq) \,   \| b_{ij}^- \|_{\Ls}   \| \chi\|_{L^{\qq}},
	\end{equation}
yielding the estimate   \eqref{S- estimate L1} from   statement of this lemma and concluding the proof.
\end{proof}

If $b_{ij} \in L^{\infty}$, then the forward and backward regions break in $\sigma$, that is $\hat{u}\cdot \sigma \geq 0 $ or $\hat{u}\cdot \sigma \leq 0$,  is not relevant anymore implying that $L^p$ norm of both $\mS^+_{ij}$ and $\mS^-_{ij}$ can be taken irrespectively in the variables $(v,I)$ or $(v_*, I_*)$. Thus, the statement of the previous lemma for bounded angular parts $b_{ij}$ can be modified as follows.

\begin{lemma}\label{Lemma S operators bdd} 
	Assume $b_{ij}\in L^{\infty}$. For $\qq\geq1$ such that the constant \eqref{cond sing r} is finite and  for a suitable test function $\chi \in L^{\qq}$, the following estimates hold on the operators defined by \eqref{S operators},
	\begin{description}
		\item[\textit{(a)}]  
		\begin{equation}\label{S+ estimate Linf}
			\sup_{(v_*, I_*)}	\left\| 	\mS^{\pm}_{ij}(\chi) \right\|_{L^{\qq}(\md v\, \md I)}  \leq 4\pi   \left( \sbar\right)^{-3/\qq} \, \cS(\qq) \,   \| b_{ij}^{\pm} \|_{\Linf}   \| \chi\|_{L^{\qq}},
		\end{equation}
		\item[\textit{(b)}]  
		\begin{equation}\label{S- estimate Linf}
			\sup_{(v, I)}	\left\| 	\mS^{\pm}_{ij}(\chi) \right\|_{L^{\qq}(\md v_*\, \md I_*)}  \leq 4\pi  \left(\sbar \right)^{-3/\qq} \, \cS(\qq) \,   \| b_{ij}^{\pm} \|_{\Linf}   \| \chi\|_{L^{\qq}}.
		\end{equation}
	\end{description}
where $\sbar$ is given in \eqref{s min}.
\end{lemma}
\begin{proof}
	The proof is a simpler version of the proof of Lemma \ref{Lemma S operators}. For the part (a), the Minkowski inequality and boundedness of  $b_{ij}$ imply
	\begin{multline*}
		\left\| 	\mS^{\pm}_{ij}(\chi) \right\|_{L^{\qq}(\md v\, \md I)}
		\\
		\leq \int_{\bS \times [0,1]^2} \left(  \int_{\bRfp} \left|   \chi(v', I') \, b_{ij}^{\pm}(\hat{u}\cdot \sigma) \, \right|^{\qq} \md v \, \md I  \right)^{1/\qq} r^{-\frac{\g}{2 \qq}} \dub \, \md r \, \md R \, \md \sigma
		\\
		\leq  \| b_{ij}^{\pm} \|_{\Linf}  \int_{\bS \times [0,1]^2} \left(    \int_{\bRfp}  \left| \chi(v', I') \right|^{\qq} \,   \md v \, \md I     \right)^{1/\qq} r^{-\frac{\g}{2 \qq}} \dub \, \md r \, \md R \, \md \sigma.
	\end{multline*}
For the 	fixed parameters $(r, R, \sigma)$,   changing   variables $(v, I) \mapsto (v', I')$ given by \eqref{p-p coll rules}, with Jacobian \eqref{Jac} implies the statement \eqref{S+ estimate Linf}.
 
The proof of part (b) is analogous.

\end{proof}

\subsection{Pointwise estimates on the velocity-internal energy part of the collision kernel}
The next goal is to provide estimates on the velocity-internal energy part $\tilde{\mathcal{B}}_{ij}$ of the collision kernel $\mathcal{B}_{ij}$ such that the weight is suitably distributed among velocity-internal energy pairs written in the bracket form. Since masses of the colliding particles can be different,   estimates require the split on two cases, when $\hat{u} \cdot \sigma \geq 0$ or $\hat{u} \cdot \sigma \leq 0$.

\begin{lemma}[Distribution of the  collision kernel]\label{Lemma weight}For the conjugate indices $p, \q \geq 1$, the following estimate on the collision kernel \eqref{p-p ass B tilde} holds
\begin{equation}\label{E decom}
\tilde{\mathcal{B}}_{ij}(v, v_*, I, I_*)	= \left( \frac{E_{ij}}{\mP}\right)^{\g/2}  \leq  \left(\tfrac{\sqrt{2}}{\sbar} \right)^{\!\!\frac{\g}{\q}}  r^{-{\frac{\g}{2\q}}}  \la v', I' \ra_i^{\frac{\g}{\q}}  \left\{ \begin{split}
&    \la v_*, I_* \ra_j^{\g}  \la v, I \ra_i^{\frac{\g}{p}}, \quad \text{for} \ \hat{u} \cdot \sigma \geq 0,\\
& \la v, I \ra_i^{\g}  \la v_*, I_* \ra_j^{\frac{\g}{p}}, \quad \text{for} \ \hat{u} \cdot \sigma \leq 0,
 	\end{split}
  \right.
\end{equation}
where  $\sbar$  is from \eqref{s min}.
\end{lemma}
\begin{proof}
On one side, 
\begin{equation}\label{E non-pr}
\sqrt{\frac{E_{ij}}{\mP}} \leq \sqrt{ \frac{m_i}{2 \, \mP} |v|^2 +  \frac{m_j}{2\, \mP} |v_*|^2 + \frac{1}{\mP} I + \frac{1}{\mP} I_* } \leq \la v, I \ra_i \la v_*, I_* \ra_j.
\end{equation}
On the other side, the aim is to incorporate primed variables. To that end, we need to consider two cases, when $\hat{u} \cdot \sigma \geq 0$ or $\hat{u} \cdot \sigma \leq 0$.

We first consider the case $\hat{u} \cdot \sigma \geq 0$. For  $v'$ from \eqref{p-p coll rules} we can write
\begin{equation*}
	v'-v_* = \frac{m_i}{m_i + m_j}  u  + \frac{m_j}{m_i + m_j} \sqrt{\frac{2 \, R  \, E_{ij}}{\mu_{ij}}} \sigma.
\end{equation*}
Taking the square, the domain yields 
\begin{equation*}
	\begin{split}
	|v'-v_*|^2 &= \left(\frac{m_i}{m_i + m_j}\right)^2  |u|^2  + \left(\frac{m_j}{m_i + m_j}\right)^2 \frac{2 \, R  \, E_{ij}}{\mu_{ij}} +   \frac{2m_i m_j}{(m_i + m_j)^2}  \sqrt{\frac{2 \, R  \, E_{ij}}{\mu_{ij}}} u \cdot \sigma
	\\
	& \geq \left(\frac{m_i}{m_i + m_j}\right)^2  |u|^2  + \left(\frac{m_j}{m_i + m_j}\right)^2 \frac{2 \, R  \, E_{ij}}{\mu_{ij}} \geq \sbar^2 \left( |u|^2  +  \frac{2 \, R  \, E_{ij}}{\mu_{ij}} \right),
	\end{split}
\end{equation*}
with $\sbar$ from \eqref{s min}. Thus, the following estimate on $|u|$ holds
\begin{equation*}
| u |^2 \leq \frac{1}{\sbar^2} |v'-v_*|^2 - \frac{2 \, R  \, E_{ij}}{\mu_{ij}}, \quad \text{for} \ \hat{u} \cdot \sigma \geq 0.
\end{equation*}
This implies for the total energy $E_{ij}$,
\begin{equation*}
	E_{ij} = \frac{\mu_{ij}}{2} | u |^2 + I + I_* \leq  \frac{\mu_{ij}}{2 \, \sbar^2} |v'-v_*|^2 - R  \, E_{ij} + I + I_*.
\end{equation*}
Moreover, since $I+I_*=(1-R')E_{ij}$ by \eqref{p-p primed param}, using \eqref{p-p coll rules}, the following estimate holds
\begin{equation}\label{I p r}
	- R  \, E_{ij} + I + I_* = (1-R-R')E_{ij} \leq (1-R)E_{ij} =  \frac{I'}{r}.
\end{equation}
Thus,
\begin{equation*}
	E_{ij} \leq   \frac{\mu_{ij}}{2 \, \sbar^2} |v'-v_*|^2  + \frac{I'}{r} + I_*.
\end{equation*}
Using the triangle inequality, 
\begin{equation*}
\frac{\mu_{ij}}{2  \, \sbar^2   } |v'-v_*|^2  \leq \frac{m_i m_j}{ (m_i+m_j)\, \sbar^2}  \left(  |v'|^2+|v_*|^2 \right) \leq  \frac{1}{  \sbar^2}   \left( {m_i} |v'|^2 +  {m_j} |v_*|^2  \right).
\end{equation*}
with $\sbar$ from \eqref{s min}.
Thus,
\begin{align}\label{E on S+}
\sqrt{\frac{E_{ij}}{\mP}} \leq \frac{\sqrt{2}}{\sbar} \frac{1}{\sqrt{r}} \sqrt{ \frac{m_i}{2 \, \mP} |v'|^2 +  \frac{m_j}{2\, \mP} |v_*|^2 + \frac{1}{\mP}I' + \frac{1}{\mP}I_* }   
 \leq  \frac{\sqrt{2}}{\sbar} \, \frac{1}{\sqrt{r}} \, \la v', I' \ra_i \, \la v_*, I_* \ra_j, \quad \text{for} \ \hat{u} \cdot \sigma \geq 0.
\end{align}
Therefore, gathering \eqref{E non-pr} and \eqref{E on S+}, the collision kernel can be estimated
\begin{equation}
	\begin{split}
	\left( \frac{E_{ij}}{\mP}\right)^{\frac{\g}{2} \left( \frac{1}{p} + \frac{1}{\q} \right)}  & \leq \Big( \la v, I \ra_i \la v_*, I_* \ra_j \Big)^{\frac{\g}{p}} \Big( \frac{\sqrt{2}}{\sbar} \, \frac{1}{\sqrt{r}} \, \la v', I' \ra_i \, \la v_*, I_* \ra_j\Big)^{\frac{\g}{\q}}
	\\
&	= \left( \tfrac{\sqrt{2}}{\sbar} \right)^{\frac{\g}{\q}} r^{-{\frac{\g}{2\q}}} \la v_*, I_* \ra_j^{\g}  \la v, I \ra_i^{\frac{\g}{p}} \la v', I' \ra_i^{\frac{\g}{\q}},
	\end{split}
\end{equation}
which proves the first part of the Lemma when $\hat{u} \cdot \sigma \geq 0$.

Similar computations can be performed for $\hat{u} \cdot \sigma \leq 0$. Indeed, involving   $v'$ from \eqref{p-p coll rules},
\begin{equation*}
v'-v =   \frac{m_j}{m_i + m_j} \left( - u  +  \sqrt{\frac{2 \, R  \, E_{ij}}{\mu_{ij}}} \sigma \right), 
\end{equation*}
and taking the square  together with the domain definition yields
\begin{equation*}
		|v'-v|^2 \geq \left(\frac{m_j}{m_i + m_j}\right)^2  \left(  |u|^2  + \frac{2 \, R  \, E_{ij}}{\mu_{ij}}   \right)\geq \sbar^2 \left(  |u|^2  + \frac{2 \, R  \, E_{ij}}{\mu_{ij}}   \right), \quad \text{for} \ \hat{u} \cdot \sigma \leq 0.
\end{equation*}
Thus, taking \eqref{I p r}, the total energy can be estimated as
\begin{equation*}
	E_{ij} \leq   \frac{\mu_{ij}}{2 \, \sbar^2  } |v'-v|^2  + \frac{I'}{r} + I.
\end{equation*}
Triangle inequality implies
\begin{equation*}
 \frac{\mu_{ij}}{2 \, \sbar^2  } |v'-v|^2  \leq 	  \frac{ m_i   }{2  \, \sbar^2 } |v'-v|^2   \leq  \frac{m_i}{  \sbar^2}    \left(  |v'|^2 +    |v|^2   \right),
\end{equation*}
 which allows to get
\begin{equation}\label{E on S-}
	\sqrt{\frac{E_{ij}}{\mP}} \leq    \frac{\sqrt{2}}{\sbar} \, \frac{1}{\sqrt{r}} \, \la v', I' \ra_i \, \la v, I \ra_i, \quad \text{for} \ \hat{u} \cdot \sigma \leq 0.
\end{equation}
This estimate \eqref{E on S-}  together with \eqref{E non-pr} yields
\begin{equation}
	\begin{split}
		\left( \frac{E_{ij}}{\mP}\right)^{\frac{\g}{2} \left( \frac{1}{p} + \frac{1}{\q} \right)}  & \leq \Big( \la v, I \ra_i \la v_*, I_* \ra_j \Big)^{\frac{\g}{p}} \Big( \frac{\sqrt{2}}{\sbar} \, \frac{1}{\sqrt{r}} \, \la v', I' \ra_i \, \la v, I \ra_i \Big)^{\frac{\g}{\q}}
		\\
		&	= \Big( \tfrac{\sqrt{2}}{\sbar} \Big)^{\frac{\g}{\q}} r^{-{\frac{\g}{2\q}}} \la v, I \ra_i^{\g}  \la v_*, I_* \ra_j^{\frac{\g}{p}} \la v', I' \ra_i^{\frac{\g}{\q}},
	\end{split}
\end{equation}
which concludes the second part of the Lemma valid for $ \hat{u} \cdot \sigma \leq 0$.
\end{proof}

\section{Estimates on the collision gain operator}\label{Sec: gain}

This section deals with the study of   gain part of pairwise collision operator $Q_{ij}^+$ from \eqref{coll gain operator}. The goal is  to provide estimates of its weak form   in a general framework using any suitable test function by applying lemmas of the previous Section \ref{Section: prelim lemmas}. The upcoming two propositions state the estimates for cases $b_{ij}\in L^1$ and $b_{ij} \in L^\infty$, respectively.

\begin{proposition}\label{Prop: Lp weak form}
	Let the gain collision operator $Q_{ij}^+$ be defined by \eqref{coll gain operator} with collision kernel $\mathcal{B}_{ij}$ satisfying assumption \eqref{p-p ass B} with $\g \in [0,2]$, whose angular part is integrable $b_{ij}\in L^1$. Then for the conjugate pair of indices $p,\q\geq1$ such that the constant $\cS(\q)$ from \eqref{cond sing r} is finite, the weight $k\geq0$ and for a suitable test function $\chi$, the following estimate holds	
	\begin{multline}\label{Lp weak form}
	\int_{\bRfp} Q_{ij}^+(f, g)(v, I) \, \chi(v, I) \,   \la v, I\ra_i^{k } \, \md v\, \md I  
	\\
	\leq 2^{1/(2\qq)} \cLp(q)  \Big(	\| b_{ij}^+ \|_{\Ls}   \| f    \|_{L^p_{i, (k+{\g})/{p}}}  \|  	g    \|_{L^1_{j, k/p+\g}} 
	+    \| b_{ij}^- \|_{\Ls}  \| 	f  \|_{L^1_{i, k/p+\g}}  \| g   \|_{L^p_{j, (k+{\g})/{p}}}    \Big)\| \chi  \|_{L^{\q}_{i, (k+ {\g})/{\q}}},
\end{multline}
where the constant  is
	\begin{equation}\label{constants Lp}
	\cLp(\qq)  = 2^{\frac{\g}{2 \qq}}  \ (\sbar)^{-\frac{3+\g}{\qq}}  \cS(\qq),
\end{equation}
with   $\sbar$ from  \eqref{s min}. 

\end{proposition}

\begin{proof} Without the loss of generality, we consider non-negative functions (otherwise consider the function's absolute value).  The proof consists of two steps. First the statement is proved for $k=0$, and then the weight $k>0$ is added.
	\subsection*{Step 1 ($k=0$)}
	Take a   suitable test function $\chi(v, I)$ and for the gain operator \eqref{coll gain operator} consider the weak form
	\begin{multline*}
		\int_{\bRfp} Q_{ij}^+(f, g)(v, I) \, \chi(v, I) \, \md v\, \md I
		\\
		=	 \int_{(\bRfp)^2} \int_{\bS \times [0,1]^2}
		f(v',I') \, g(v'_*,I'_*) \left(\frac{I }{I' }\right)^{\alpha_i } \left(\frac{ I_*}{ I'_*}\right)^{ \alpha_j}   \chi(v,I)  \\ \times \mathcal{B}_{ij} \, \dpo  \, \md \sigma \, \md r \, \md R \, \md v_* \, \md I_*  \, \md v \, \md I. 
	\end{multline*}
	We interchange pre- and post-collisional quantities, i.e. we perform the change of variables \eqref{pre-post T}. Recalling the Jacobian of transformation \eqref{p-p Jac}, together with the invariance properties \eqref{p-p Bij micro} and \eqref{meas inv}, the last equation becomes 
	\begin{multline*}
		\int_{\bRfp} Q_{ij}^+(f, g)(v, I) \, \chi(v, I) \, \md v\, \md I
		\\
		=	 \int_{(\bRfp)^2} \int_{\bS \times [0,1]^2}
		f(v,I) \, g(v_*,I_*) \,  \chi(v',I') \, \mathcal{B}_{ij} \, \dpo  \, \md \sigma \, \md r \, \md R \, \md v_* \, \md I_*  \, \md v \, \md I. 
	\end{multline*}
Using the assumption on  collision kernel \eqref{p-p ass B}, the last equation gives
  \begin{multline}\label{Q chi}
	\int_{\bRfp} Q_{ij}^+(f, g)(v, I) \, \chi(v, I) \, \md v\, \md I
	\\
%	=	 \int_{(\bRfp)^2} \int_{\bS \times [0,1]^2}
%	f(v,I) \, g(v_*,I_*) \,  \chi(v',I') \, \mathcal{B}_{ij} \, \dpo  \, \md \sigma \, \md r \, \md R \, \md v_* \, \md I_*  \, \md v \, \md I
%	\\ 
	\leq \int_{(\bRfp)^2} \int_{\bS \times [0,1]^2}
	f(v,I) \, g(v_*,I_*)\, \tilde{\mathcal{B}}_{ij}(v,I,v_*,I_*) 
	\\ \times \chi(v',I') \, b_{ij}(\hat{u}\cdot\sigma) \, \dub  \, \md \sigma \, \md r \, \md R \, \md v_* \, \md I_*  \, \md v \, \md I =: T_1 + T_2.
\end{multline}
Now we use the form of $ \tilde{\mathcal{B}}_{ij}$ given in \eqref{p-p ass B tilde} and its estimate provided in \eqref{E decom}. Thus, it is necessary to consider separately   cases $\hat{u}\cdot\sigma \geq 0$ and $\hat{u}\cdot\sigma \leq 0$ and terms $T_1$ and $T_2$ corresponding respectively to these cases.

\subsubsection*{Term $T_1$}
When  $\hat{u}\cdot \sigma\geq 0$, the corresponding term $T_1$ can be estimated using the decomposition of  velocity-internal energy part $\tilde{\mathcal{B}}_{ij}$ of the collision kernel given in \eqref{E decom} and rewritten in terms of $\mS^+_{ij}$ defined in \eqref{S operators}, namely
\begin{multline*}
	T_1 := \int_{(\bRfp)^2} \int_{\bS \times [0,1]^2}
	f(v,I) \,  g(v_*,I_*)\, \tilde{\mathcal{B}}_{ij}(v,I,v_*,I_*) 
	\\ \times \chi(v',I') \, b^+_{ij}(\hat{u}\cdot\sigma) \, \dub  \, \md \sigma \, \md r \, \md R \, \md v_* \, \md I_*  \, \md v \, \md I
	\\
\leq   \left(\tfrac{\sqrt{2}}{\sbar}  \right)^{\frac{\g}{\q}}	 \int_{(\bRfp)^2} \int_{\bS \times [0,1]^2}
	f(v,I) \, \la v, I \ra_i^{\frac{\g}{p}}\,  g(v_*,I_*)\,  \la v_*, I_* \ra_j^{\g} 
	\\ \times \chi(v',I') \,  \la v', I' \ra_i^{\frac{\g}{\q}}  \, b^+_{ij}(\hat{u}\cdot\sigma) \, r^{-{\frac{\g}{2\q}}}  \dub  \, \md \sigma \, \md r \, \md R \, \md v_* \, \md I_*  \, \md v \, \md I
\end{multline*}
With the operator $	\mS^+_{ij}$ defined as in \eqref{S operators}, we can rewrite the last inequality, 
\begin{equation}\label{proof Lp weak form +}
T_1 \leq  \left(\tfrac{\sqrt{2}}{\sbar}  \right)^{\frac{\g}{\q}}	 \int_{(\bRfp)^2} 
	f(v,I) \, \la v, I \ra_i^{\frac{\g}{p}}\,  g(v_*,I_*)\,  \la v_*, I_* \ra_j^{\g} 	\mS^+_{ij}(\chi \la \cdot \ra_i^{{\g}/{\q}}  )(v, I, v_*, I_*)\,
  \md v_* \, \md I_*  \, \md v \, \md I.
\end{equation}
Then H\"older's inequality applied to the integral with respect to $(v,I)$ yields, with notation \eqref{Lp (v,I)},
\begin{equation*}
	T_1 \leq \left(\tfrac{\sqrt{2}}{\sbar}  \right)^{\frac{\g}{\q}}	 \| f  \la \cdot \ra_i^{ {\g}/{p}}  \|_{L^p}  \int_{\bRfp} 
 g(v_*,I_*)\,  \la v_*, I_* \ra_j^{\g}  \|	\mS^+_{ij}(\chi \la \cdot \ra_i^{{\g}/{\q}})\|_{L^{\q}(\md v\, \md I)} 
	\md v_* \, \md I_*.
\end{equation*} 
Using the estimate \eqref{S+ estimate L1} from Lemma \ref{Lemma S operators}, one has
\begin{equation}\label{T1}
	\begin{split}
	T_1 &\leq \left(\tfrac{\sqrt{2}}{\sbar}  \right)^{\frac{\g}{\q}} 2^{1/(2\qq)}  \left( \frac{m_i}{m_i+m_j} \right)^{-3/\qq} \, \cS(\qq) \,	\| b_{ij}^+ \|_{\Ls} \| \chi \la \cdot \ra_i^{{\g}/{\q}} \|_{L^{\qq}}  \| f  \la \cdot \ra_i^{ {\g}/{p}}  \|_{L^p}  \|  	g \, \la \cdot \ra_j^{\g}      \|_{L^1} 
	\\
	&= 2^{1/(2\qq)} \,  \cLp \,	\| b_{ij}^+ \|_{\Ls} \| \chi \la \cdot \ra_i^{{\g}/{\q}} \|_{L^{\qq}}  \| f  \la \cdot \ra_i^{ {\g}/{p}}  \|_{L^p}  \|  	g \, \la \cdot \ra_j^{\g}      \|_{L^1},
	\end{split}
\end{equation} 
with notation \eqref{constants Lp}.

\subsubsection*{Term $T_2$}
On the other hand, when $\hat{u}\cdot \sigma\leq 0$, with the estimate \eqref{E decom}, term $T_2$ reads
\begin{multline}\label{proof Lp weak form -}
	T_2 := \int_{(\bRfp)^2} \int_{\bS \times [0,1]^2}
	f(v,I) \,  g(v_*,I_*)\, \tilde{\mathcal{B}}_{ij}(v,I,v_*,I_*) 
	\\ \times \chi(v',I') \, b^-_{ij}(\hat{u}\cdot\sigma) \, \dub  \, \md \sigma \, \md r \, \md R \, \md v_* \, \md I_*  \, \md v \, \md I
	\\
	\leq   \left(\tfrac{\sqrt{2}}{\sbar}  \right)^{\frac{\g}{\q}} \int_{(\bRfp)^2} 
	f(v,I) \,  \la v, I \ra_i^{\g} \,  g(v_*,I_*)\, \la v_*, I_* \ra_j^{\frac{\g}{p}} 
	\mS^-_{ij}(\chi \la \cdot \ra_i^{{\g}/{\q}}  )(v, I, v_*, I_*)\,
\md v_* \, \md I_*  \, \md v \, \md I.
\end{multline}
Following the same strategy as for $T_1$ and applying the H\"older inequality for $(v_*, I_*)$ integration in the present case, together with Lemma \ref{Lemma S operators} and the estimate \eqref{S- estimate L1}, implies
\begin{equation}\label{T2}
\begin{split}	
	T_2 & \leq   \left(\tfrac{\sqrt{2}}{\sbar} \right)^{\frac{\g}{\q}}  \| g \, \la \cdot \ra_j^{ {\g}/{p}}  \|_{L^p}  \int_{\bRfp} 
	f(v,I) \,  \la v, I \ra_i^{\g} \,  
\|	\mS^-_{ij}(\chi \la \cdot \ra_i^{{\g}/{\q}})\|_{L^q(v_*, I_*)} \, \md v \, \md I
\\ 
&\leq 2^{1/(2\qq)}   \cLp \,  \| b_{ij}^- \|_{\Ls}   \| \chi   \la \cdot \ra_i^{{\g}/{\q}}\|_{L^{\qq}} \| g \, \la \cdot \ra_j^{ {\g}/{p}}  \|_{L^p}   \| 	f \,  \la \cdot \ra_i^{\g} \|_{L^1},
\end{split}
\end{equation}
with the constant \eqref{constants Lp}.

\medspace

Gathering estimates \eqref{T1} and  \eqref{T2} for $T_1$ and $T_2$, \eqref{Q chi} becomes
\begin{align}\label{result Step 1}
	&	\int_{\bRfp}  Q_{ij}^+(f, g)(v, I) \, \chi(v, I) \, \md v\, \md I \nonumber
		\\
	&	\leq 2^{1/(2\qq)}  \, \cLp  \left( 	\| b_{ij}^+ \|_{\Ls}   \| f  \la \cdot \ra_i^{ {\g}/{p}}  \|_{L^p}  \|  	g \, \la \cdot \ra_j^{\g}      \|_{L^1} 
		+   \| b_{ij}^- \|_{\Ls}  \| 	f \,  \la \cdot \ra_i^{\g} \|_{L^1}  \| g \, \la \cdot \ra_j^{ {\g}/{p}}  \|_{L^p}    \right)\| \chi   \la \cdot \ra_i^{{\g}/{\q}}\|_{L^{\qq}} \nonumber
			\\
	&	= 2^{1/(2\qq)}  \, \cLp \,  \left( 	\| b_{ij}^+ \|_{\Ls}   \| f    \|_{L^p_{i,  {\g}/{p}}}  \|  	g    \|_{L^1_{j,\g}} 
		+     \| b_{ij}^- \|_{\Ls}  \| 	f  \|_{L^1_{i, \g}}  \| g   \|_{L^p_{j, {\g}/{p}}}    \right)\| \chi  \|_{L^{\q}_{i,{\g}/{\q}}},
\end{align}
with notation of functional spaces introduced in \eqref{Lp i}.
\subsection*{Step 2 ($k>0$)} 
From the collisional energy conservation law \eqref{p-p coll CL}, it follows
\begin{equation*}
\la v, I \ra_i^2 = 1+ \frac{m_i}{2 \,\mP} |v|^2 + \frac{1}{\mP} I \leq 1+ 	\frac{m_i}{2 \, \mP} |v'|^2 + \frac{1}{\mP} I' +  \frac{m_j }{2 \, \mP} |v'_*|^2 + \frac{1}{\mP} I'_* \leq \la v', I' \ra_i^2 \,\la v'_*, I'_* \ra_j^2.
\end{equation*}
Thus, the strong form of the collision operator \eqref{coll gain operator} can be estimated, for any $\ell \geq0$,
\begin{multline*}
	Q^+_{ij}(f,g)(v,I) \, \la v, I\ra_i^{\ell} 
	\\
	\leq \int_{\bRfp} \int_{\bS \times [0,1]^2}   f(v',I') \, \la v', I'\ra_i^{\ell} \, g(v'_*,I'_*) \, \la v'_*, I'_*\ra_j^{\ell}  \left(\frac{I }{I' }\right)^{\alpha_i } \left(\frac{ I_*}{ I'_*}\right)^{ \alpha_j}      \mathcal{B}_{ij}  \, d_{ij}  \, \md \sigma \, \md r \, \md R \, \md v_* \, \md I_*
	\\ 	
	= Q^+_{ij}(f\la \cdot \ra_i^{\ell}, g\la \cdot \ra_j^{\ell})(v,I).
\end{multline*}
Finally, the weight is rewritten
\begin{equation*}
 \la v, I\ra_i^{k  \left(\frac{1}{p} + \frac{1}{\q}\right)}  =  \la v, I\ra_i^{ {k}/{p}} \  \la v, I\ra_i^{ {k}/{\q}},
\end{equation*}
so that it can be distributed to the collision operator and to the test function, namely
\begin{multline*}
	\int_{\bRfp} Q_{ij}^+(f, g)(v, I) \, \chi(v, I) \,   \la v, I\ra_i^{k } \, \md v\, \md I  
	\\ = 	\int_{\bRfp} Q_{ij}^+\left(f\la \cdot \ra_i^{{k}/{p}}, g\la \cdot \ra_j^{{k}/{p} }\right)\!(v,I) \, \chi(v, I) \,   \la v, I\ra_i^{ {k}/{\q}} \, \md v\, \md I
	\\ = 	\int_{\bRfp} Q_{ij}^+(\tilde{f}_i, \tilde{g}_j)(v,I) \,  \omega_i(v, I)  \, \md v\, \md I,
\end{multline*}
with notation
\begin{equation*}
\tilde{f}_i(v, I) = f(v, I) \la v, I \ra_i^{k/p}, \quad \tilde{g}_j(v, I) = g(v, I)\la v, I \ra_j^{k/p }, \quad \omega_i(v, I) =  \chi(v, I) \,   \la v, I \ra_i^{{k}/{\q}}.
\end{equation*}
Then we simply apply the result  \eqref{result Step 1} for $k=0$ and get
\begin{multline*}
	\int_{\bRfp}Q_{ij}^+(\tilde{f}_i, \tilde{g}_j)(v,I) \,  \omega_i(v, I)  \, \md v\, \md I 
	\\
	\leq 2^{1/(2\qq)}  \, \cLp \left( 	\| b_{ij}^+ \|_{\Ls}   \| \tilde{f}_i    \|_{L^p_{i, {\g}/{p}}}  \|  	\tilde{g}_j    \|_{L^1_{j, \g}} 
	+     \| b_{ij}^- \|_{\Ls}  \| 	\tilde{f}_i  \|_{L^1_{i,\g}}  \| \tilde{g}_j   \|_{L^p_{j, {\g}/{p}}}    \right)\| \omega_i  \|_{L^{\q}_{i, {\g}/{\q}}}
		\\
	=  2^{1/(2\qq)}  \, \cLp  \left( 	\| b_{ij}^+ \|_{\Ls}   \| f    \|_{L^p_{i, (k+{\g})/{p}}}  \|  	g    \|_{L^1_{j, \, k/p+\g}} 
	+    \| b_{ij}^- \|_{\Ls}  \| 	f  \|_{L^1_{i, \,k/p+\g}}  \| g   \|_{L^p_{j, (k+{\g})/{p}}}    \right)\| \chi  \|_{L^{\q}_{i,(k+ {\g})/{\q}}},
\end{multline*}
which is exactly the desired estimate \eqref{Lp weak form}.

\end{proof}

Note that in the estimate \eqref{Lp weak form} for integrable angular part of the collision kernel the order of the input functions in $Q_{ij}^+$ plays a role since the $L^p$ norms of the input functions are carefully combined with either  $b_{ij}^+$ or  $b_{ij}^-$.   As stated in Lemma \ref{Lemma S operators}, the reason lies in the integrability constraint of the $\mS^{\pm}_{ij}$ operators raising from the collision's lack of symmetry. Namely, the forward scattering  $b_{ij}^+$ can only produce $L^p$ norm of the first entry function $f$ and back scattering $b_{ij}^-$ is related to the $L^p$ norm of the second entry function $g$. 

When the angular kernel is assumed bounded, the result of the aforementioned Proposition \ref{Prop: Lp weak form} can be applied. More precisely, the scattering in each direction, either forward or backward, can be controled by the $L^p$ norm of any of the input functions, albeit with a different polynomial order, due to Lemma \ref{Lemma weight}.

\begin{proposition}\label{Prop: Lp weak form bdd}
	For the collision operator gain part $Q_{ij}^+$ given in \eqref{coll gain operator} with the collision kernel $\mathcal{B}_{ij}$ satisfying assumption \eqref{p-p ass B} with $\g \in[0,2]$ and $b_{ij} \in L^\infty$  decomposed as in \eqref{bij decom}, conjugate indices $p, q \geq1$ such that the constant $\cS(\q)$ from \eqref{cond sing r} is finite,  a factor $k\geq 0$ and a suitable test function $\chi$, the following estimates hold
\begin{description}
	\item[(a)] for $b^+_{ij} $ having the support in the set $\left\{ \hat{u} \cdot \sigma \geq0 \right\}$,
\begin{multline}\label{Lp weak form inf +}
	\int_{\bRfp} Q_{ij}^+(f, g)(v, I) \, \chi(v, I) \,   \la v, I\ra_i^{k } \, \md v\, \md I  
	\\
	\leq   4\pi \| b_{ij}^+ \|_{\Linf}  \,  \cLp(q)\,  	\| \chi  \|_{L^{\q}_{i, (k+ {\g})/{\q}}} 
	\begin{cases}
		&  \| f    \|_{L^p_{i, (k+{\g})/{p}}}  \|  	g    \|_{L^1_{j, k/p+\g}},\\
		&    \| g    \|_{L^p_{j, k/{p}+\g}}  \|  	f    \|_{L^1_{i, (k+{\g})/{p}}},
	\end{cases} 
\end{multline}
	\item[(b)]  for $b^-_{ij} $ having the support in the set $\left\{ \hat{u} \cdot \sigma \leq0 \right\}$,
\begin{multline}\label{Lp weak form inf -}
\int_{\bRfp} Q_{ij}^+(f, g)(v, I) \, \chi(v, I) \,   \la v, I\ra_i^{k } \, \md v\, \md I  
\\
\leq	  4\pi \| b_{ij}^- \|_{\Linf} \, \cLp(q) \,  	\| \chi  \|_{L^{\q}_{i, (k+ {\g})/{\q}}} 
\begin{cases}
	&    \| g   \|_{L^p_{j, (k+{\g})/{p}}}  \|  	f   \|_{L^1_{i, k/p+\g}},\\
	&        \| f    \|_{L^p_{i, k/p+{\g}}}  \|  	g    \|_{L^1_{j, (k+\g)/p}},
\end{cases} 
\end{multline}
\end{description}
with the constant $\cLp(q)$ from \eqref{constants Lp}.
\end{proposition}

\begin{proof}
Consider the angular part of the collision kernel decomposed as in \eqref{bij decom} and    first consider the collision operator $Q_{ij}^+$ for $b_{ij}^+$. The first part of the statement   \eqref{Lp weak form inf +}$_1$ corresponds to the one for $b_{ij} \in L^1$ from Proposition \ref{Prop: Lp weak form}, and it is a matter of estimating the norm 
\begin{equation*}
	\| b_{ij}^+ \|_{\Ls} \leq 4\pi 	\| b_{ij}^+ \|_{\Linf},
\end{equation*}
and using the appropriate constant from Lemma \ref{Lemma S operators bdd}.
For the second part  \eqref{Lp weak form inf +}$_2$ , taking $k=0$ and proceeding in the same way as in the proof of Proposition \ref{Prop: Lp weak form}, for the starting point here we take the estimate \eqref{proof Lp weak form +},
	\begin{multline}\label{pomocna 6}
\int_{\bRfp} Q_{ij}^+(f, g)(v, I) \, \chi(v, I)  \, \md v\, \md I 
\\
  \leq  \left(\tfrac{\sqrt{2}}{\sbar}  \right)^{\frac{\g}{\q}}	 \int_{(\bRfp)^2} 
		f(v,I) \, \la v, I \ra_i^{\frac{\g}{p}}\,  g(v_*,I_*)\,  \la v_*, I_* \ra_j^{\g} 	\mS^+_{ij}(\chi \la \cdot \ra_i^{{\g}/{\q}}  )(v, I, v_*, I_*)\,
		\md v_* \, \md I_*  \, \md v \, \md I.
	\end{multline}
Contrary to the case of $b_{ij} \in L^1$,       for $b_{ij} \in L^\infty$   the estimate \eqref{S- estimate Linf} is valid also for 		$\mS^+_{ij}$, implying that  H\"older's inequality can be applied for integration with respect to $(v_*, I_*)$, i.e. for any $(v,I)$,
\begin{multline*}
 \int_{\bRfp} 
   g(v_*,I_*)\,  \la v_*, I_* \ra_j^{\g} 	\mS^+_{ij}(\chi \la \cdot \ra_i^{{\g}/{\q}}  )(v, I, v_*, I_*)\,
\md v_* \, \md I_* 
\\
 \leq \| g \la \cdot \ra_j^{\g} \|_{L^p} \| \mS^+_{ij}(\chi \la \cdot \ra_i^{{\g}/{\q}}  ) \|_{L^q(\md v_* \md I_*)}
%\\
%\leq 4\pi  \left( \frac{m_j}{m_i+m_j} \right)^{-3/\qq} \, \cS(\qq) \,   \| b_{ij}^{+} \|_{\Linf}   \| g \la \cdot \ra_j^{\g} \|_{L^p}   \| \chi \la \cdot \ra_i^{{\g}/{\q}}  \|_{L^{\qq}}
%\\
\leq 4\pi  \left( \sbar \right)^{-3/\qq} \, \cS(\qq) \,   \| b_{ij}^{+} \|_{\Linf}   \| g   \|_{L^p_{j, \g}}   \| \chi   \|_{L^{\qq}_{i, \g/q}}.
\end{multline*}
Returning this result into \eqref{pomocna 6} leads to the second statement \eqref{Lp weak form inf +}$_2$, after the weight factor $k>0$ is added similarly as in the second step of the proof of Proposition \ref{Prop: Lp weak form}. \\
 
Next, consider the  gain operator $Q_{ij}^+$ for the  angular part of the collision kernel $b_{ij}^-$. As in the previous case, the first statement \eqref{Lp weak form inf -}$_1$ is analogous to the corresponding  statement of Proposition \ref{Prop: Lp weak form}. To prove the second part of the statement \eqref{Lp weak form inf -}$_2$, start from \eqref{proof Lp weak form -},
 \begin{multline}\label{pomocna 7}
 \int_{\bRfp} Q_{ij}^+(f, g)(v, I) \, \chi(v, I)  \, \md v\, \md I 
 	\\
 	\leq   \left(\tfrac{\sqrt{2}}{\sbar}  \right)^{\frac{\g}{\q}} \int_{(\bRfp)^2} 
 	f(v,I) \,  \la v, I \ra_i^{\g} \,  g(v_*,I_*)\, \la v_*, I_* \ra_j^{\frac{\g}{p}} 
 	\mS^-_{ij}(\chi \la \cdot \ra_i^{{\g}/{\q}}  )(v, I, v_*, I_*)\,
 	\md v_* \, \md I_*  \, \md v \, \md I.
 \end{multline}
 The H\"older inequality for integration with respect to $(v, I)$,  together with the estimate on 	$\mS^-_{ij}$ operator given in \eqref{S+ estimate Linf}, yields for any $(v_*, I_*)$,
 \begin{multline*}
 	\int_{\bRfp} 
 	f(v,I) \,  \la v, I \ra_i^{\g} \,   
 \mS^-_{ij}(\chi \la \cdot \ra_i^{{\g}/{\q}}  )(v, I, v_*, I_*) \, \md v \, \md I
 	\\
 	\leq \| f \la \cdot \ra_i^{\g} \|_{L^p} \| \mS^-_{ij}(\chi \la \cdot \ra_i^{{\g}/{\q}}  ) \|_{L^q(\md v \, \md I)}
 	%\\
 	%\leq 4\pi  \left( \frac{m_j}{m_i+m_j} \right)^{-3/\qq} \, \cS(\qq) \,   \| b_{ij}^{+} \|_{\Linf}   \| g \la \cdot \ra_j^{\g} \|_{L^p}   \| \chi \la \cdot \ra_i^{{\g}/{\q}}  \|_{L^{\qq}}
 	%\\
 	\leq 4\pi  \left( \sbar \right)^{-3/\qq} \, \cS(\qq) \,   \| b_{ij}^{-} \|_{\Linf}   \| f   \|_{L^p_{i, \g}}   \| \chi   \|_{L^{\qq}_{i, \g/q}}.
 \end{multline*}
 Plugging this result into \eqref{pomocna 7} and adding weight factor $k>0$ yields \eqref{Lp weak form inf -}$_2$.

\end{proof}

\begin{remark}[Discussion of the integrability condition on  $\cS$ when  $\tilde{b}_{ij}^{ub}$ is constant ]\label{Remark const S op}
In this remark, we will discuss integrability conditions for the constant $\cS(q)$ defined in \eqref{cond sing r} for a particular case of the collision kernel assumption \eqref{p-p ass B} when $\tilde{b}_{ij}^{ub} = 1$ which, for instance, includes the choice of the collision kernel \eqref{Bij choice energy}. Namely, the following conditions relating parameters  $\g \in [0,2]$, $\alpha_i , \alpha_j>-1$  and $\qq\geq 1$ have to be satisfied in order to guarantee finiteness of     $\cS(q)$,
	\begin{equation*}
\frac{1}{q} < \frac{(1+\alpha_i)}{(1+\g/2)}, \qquad \frac{1}{q} < \alpha_i + \alpha_j +2,
	\end{equation*}
implying for $p$ the following conditions
\begin{equation}\label{p restr}
	\left( \frac{\g}{2} - \alpha_i \right) p < 1+\frac{\g}{2 }, \quad p \, (1+\alpha_i+\alpha_j) > -1.
\end{equation}
The aforementioned conditions restrict the values of $p$ in two particular cases, namely,
\begin{description}
	\item[\textit{Case (i)}]  if $\alpha_i  < \g/2$, then \eqref{p restr} gives the upper bound on the value of $p$,
	\begin{equation}\label{p restr c1}
		p <   \frac{1+\g/2}{\g/2-\alpha_i},
	\end{equation}
\item[\textit{Case (ii)}]  if $\alpha_i+\alpha_j< -1 $, then \eqref{p restr}  implies the upper bound on $p$ as follows
\begin{equation}\label{p restr c2}
  p<-\frac{1}{1+\alpha_i + \alpha_j}.
\end{equation}
\end{description}
Although the model of  collision operator  \eqref{p-p coll operator} and   $H$-theorem \cite{DesMonSalv, LD-ch-ens} allow parameters $\alpha_i, \alpha_j$ related to a structure of polyatomic molecules to have values strictly greater than $-1$,  in applications they are rarely negative.  For instance, the typical rough estimate of $\alpha_i$ corresponding to the modelling of linear and non-linear molecules at room temperature is $\alpha_i =0 $ and $\alpha_i=0.5$ respectively.  More accurate values of $\alpha_i$ can be obtained using the gases specific heats experimental data \cite{MPC-SS-non-poly, MPC-Dj-T-O}.  In both interpretations, $\alpha_i$ does not satisfy the  condition of the Case (ii), namely $\alpha_i+\alpha_j< -1 $,   regardless the value of $\alpha_j>-1$, and thus \eqref{p restr c2} essentially does not restrict the value of $p$ in applications.

%\begin{figure}
%	\includegraphics{p-values.pdf}
%	\caption{The upper limiting curve \eqref{p restr c1} for the value of $p$ as a function of $\gamma_{ij} \in (0,2]$ for the particular choice of $\alpha_i=0$ and $\alpha_i=0.5$ corresponding to rough estimates of this parameter when linear and non-linear molecules, respectively,  at room temperature are modelled. The choice $\alpha=0$ leads to the restriction on the whole range $\g \in (0,2]$, whereas $\alpha=0.5$ restricts the value of $p$ only in the region $\g \in (1,2]$. }
%		\label{Fig 1}
%\end{figure}

The Case (i) gives the upper limit for the value of $p$ for a specific combination of $\alpha_i$ and $\g$.  For an illustration let us discuss two typical choices of $\alpha_i=0$ and $\alpha_i = 0.5$. The choice $\alpha_i=0$ gives the upper limiting curve for $p$ on the whole range of $\g \in [0,2]$ and the lowest value is  $p=2$ attained when $\g=2$. For  $\alpha_i = 0.5$ condition \eqref{p restr c1} restricts the value of $p$ only in the region $\gamma_{ij} \in (1,2]$; in $\g \in (0,1]$ this condition is immaterial and $p \geq1$ can take any value.  In applications the non-restrictive  value $\g=1$  is used in the shock structure problem \cite{MPC-SS-Madj-Symm}, lately exploited to obtain estimates of phenomenological coefficients for instance in \cite{Rugg-shock}.

%  Moreover, $\gamma_{ij}$ can be related to measurements as well. Namely, it has a physical interpretation of  the  shear viscosity exponent and thus can be recovered from experiments  \cite{MPC-SS-non-poly, MPC-Dj-T-O}. Table \ref{Tab 1: poly gas} lists the value of $\g$ that combined with the value of $\alpha_i$ gives the limiting value of $p$ according to \eqref{p restr c1}.

%
% 
% 	\begin{table}
% 	\begin{tabular}{|c | c | c | c | c | c | c |}\hline
% 		gas & N$_2$ & O$_2$ & NO & CO  &  H$_2$ \\ \hline 
% 		$\alpha_i$  & 0.0023 & 0.0348 & 0.0901 & 0.0059  & -0.0301 \\ \hline
% 		$\gamma_{ii}$ & 0.533 & 0.441  & 0.42 & 0.53 & 0.607 \\ \hline
% 		p & 4.79 & 6.57 & 10.09 & 4.88 & 3.91 \\ \hline
% 	\end{tabular}
% 	\caption{ Values of $\alpha_i$ and $\gamma_{ii}$   obtained by matching to experimental data  and the corresponding limiting  value of $p$ valid for $\alpha_i  < \g/2$, and computed from  the integrability condition \eqref{p restr c1} of  $\cS$ when  $\tilde{b}_{ij}^{ub}$ is constant. } 
% 	\label{Tab 1: poly gas}
% \end{table}
% 

\end{remark}

\section{Collision frequency lower bound}\label{Sec: loss}

This section deals with finding lower bounds of the collision frequency defined in \eqref{coll fr} that will immediately lead to a lower bound of the collision loss term. First, we prove the lower bound of the weak form related to the velocity-internal energy part \eqref{p-p ass B tilde} of the collision kernel combined with a sufficiently integrable function $g$ having the following entropy-like quantity finite
\begin{equation}\label{entropy H0}
	H[g]  :=    \int_{\bRfp}       g(v, I)  \left| \log g(v, I) \right|   \,\md v \, \md I < \infty.
\end{equation}
This quantity can be related to the initial entropy for solutions of \eqref{BE vector}.  Indeed, let
%Before entering in details consider 
$\F =   \Big[f_i(t, v, I)\Big]_{i=1,\dots,P}$ be a solution of \eqref{BE vector} with finite initial entropy
\begin{equation*}
\sum^{P}_{i=1}\int_{\bRfp}  f_{0,i}(v, I)  \log f_{0,i}(v, I)    \,\md v \, \md I = \mathcal{H}(\F_0)< \infty\,.
\end{equation*}
Using the elementary inequality $x|\log(x)|\leq x\log(x) + 2x^{\frac34}$ and the system dissipation of entropy it follows that, with some abuse of notation allowing $H[\cdot]$ to have vector or scalar entries,
\begin{multline*}
H[\F] := \sum^{P}_{i=1} H[f_i] = \sum^{P}_{i=1} \int_{\bRfp}       f_{i}(t,v, I)  \left| \log f_{i}(t,v, I) \right|   \,\md v \, \md I 
\\
\leq \sum^{P}_{i=1}\int_{\bRfp}  f_{i}(t,v, I)  \log f_{i}(t,v, I)    \,\md v \, \md I 
+ 2\sum^{P}_{i=1}\int_{\bRfp}  \big(f_{i}(t,v, I)\big)^{\frac34}   \,\md v \, \md I
\\
\leq \mathcal{H}(\F_0) + 2\sum^{P}_{i=1}\int_{\bRfp}  \big(f_{i}(t,v, I)\big)^{\frac34}   \,\md v \, \md I\,.
\end{multline*}
Additionally, using H\"{o}lder's inequality
\begin{equation*}
\int_{\bRfp}  \big(f_{i}(t,v, I)\big)^{\frac34}   \,\md v \, \md I \leq \| f_i \|^{\frac34}_{L^{1}_{i,2}} \| \langle \cdot \rangle^{-6}_{i}  \|^{\frac14}_{L^{1}_0} \,,
\end{equation*}
and consequently the following estimate is obtained for the vector valued solution $\F$,
\begin{equation}\label{H finite}
H[\F]  \leq \mathcal{H}(\F_0) + 2P^{\frac14}\max_{i}\| \langle \cdot \rangle^{-6}_{i} \|^{\frac14}_{L^{1}_0}  \| \F_0 \|^{\frac34}_{L^{1}_{2}}\,.
\end{equation} 
Thus, we conclude that solutions $\F$ of \eqref{BE vector} with finite initial entropy satisfy $\sup_{t\geq0}H[\F] < \infty$ with bound depending only on the naturally propagated quantities of the mixture mass, energy and entropy.
 
\begin{lemma}[Lower Bound Lemma] \label{Lemma Low b}Assume $g \in {L^1_{j, \g}}$,  $\g \in[0,2]$, with finite $H[g] $ defined in \eqref{entropy H0}. 
Then, there exists a constant $\tilde{c}^{lb}_{ij}[g]  > 0$ that depends on $H[g]$ and on $ \|g\|_{L^1_{j, \g}}$, explicitly computed in \eqref{c lower bound}, such that for the velocity-internal energy part $ \tilde{\mathcal{B}}_{ij}$ of the collision kernel  \eqref{p-p ass B tilde} the following estimate holds
\begin{equation}\label{lbl}
	\int_{\bRfp}  g(v_*, I_*) \, \tilde{\mathcal{B}}_{ij}(v, v_*, I, I_*) \, \md v_* \, \md I_* \geq   \tilde{c}^{lb}_{ij}[g] \, \la v, I \ra_{i}^{\g}.
\end{equation}
\end{lemma}
\begin{proof} The proof is divided into three steps.
\subsection*{Step 1} 
Triangle inequality $|v-v_*| \geq |v| - |v_*|$ and few rearrangements imply
\begin{equation*}
	\begin{split}
		\sqrt{\frac{E_{ij}}{\mP}} & \geq \frac{1}{\sqrt{2}}  \left( \sqrt{\frac{\mu_{ij}}{ 2\mP}}  |v-v_*| + \sqrt{\frac{I}{\mP}} \right) \geq \sqrt{ \frac{m_j}{2(m_i+m_j)} } \left(  \sqrt{\frac{m_i}{2 \mP}} |v|    + \sqrt{\frac{I}{\mP}} \right) - \sqrt{\frac{m_j}{2 \mP}} |v_*|  \\
		&\geq \sqrt{ \frac{m_j}{2(m_i+m_j)} }  \sqrt{\frac{m_i}{2 \mP} |v|^2  + \frac{I}{\mP}  }  -   \la v_*, I_* \ra_j
 	\geq \sqrt{\frac{\sbar}{2}}   \sqrt{\frac{m_i}{2 \mP} |v|^2  + \frac{I}{\mP}  }  -   \la v_*, I_* \ra_j.
	\end{split}
\end{equation*}
Raising to the power $\g  \in [0,2]$, 
\begin{equation*}
	 \left( \frac{\sbar}{2}\right)^{\!\!\frac{\g}{2}}  	 \left( \frac{m_i}{2 \mP} |v|^2  + \frac{I}{\mP} \right)^{\!\!\frac{\g}{2}}  \leq \max\{1, 2^{\g-1}\} \left(  \tilde{\mathcal{B}}_{ij} + \la v_*, I_*\ra_j^{\g} \right).
\end{equation*} 
Therefore,  $ \tilde{\mathcal{B}}_{ij}$ can be  estimated pointwise from below as 
\begin{equation*}
\tilde{\mathcal{B}}_{ij} \geq L_1	 \left( \frac{m_i}{2 \mP} |v|^2  + \frac{I}{\mP} \right)^{\!\!\frac{\g}{2}}  - \la v_*, I_*\ra_j^{\g}, \quad \text{with} \quad L_1= \min\{1, 2^{1-\g}\}	\left( \frac{\sbar}{2}\right)^{\!\!\frac{\g}{2}},
\end{equation*} 
leading to
\begin{equation}
	\label{lbl key 1}
\int_{\bRfp}  g(v_*, I_*) \, \tilde{\mathcal{B}}_{ij} \, \md v_* \, \md I_* \geq L_1 \, \|g\|_{L^1_0} 	 \left( \frac{m_i}{2 \mP} |v|^2  + \frac{I}{\mP} \right)^{\!\!\frac{\g}{2}} - \|g\|_{L^1_{j, \g}}.
\end{equation}

 \subsection*{Step 2} 
  For $\eta>0$, define the set
  \begin{equation*}
  	B_\eta(v, I) = \left\{ (v_*, I_*) \in \bRfp: \sqrt{\frac{E_{ij}}{\mP}} = \sqrt{ \frac{\mu_{ij}}{2 \mP} |v-v_*|^2 + \frac{I+I_*}{\mP}} \leq \eta \right\},
  \end{equation*}
and with $B_\eta^c$ we denote its complement.

 Then for some $\delta>0$ to be chosen later, 
 \begin{equation}\label{pomocna 3}
 	\begin{split}
 	\int_{\bRfp}  g(v_*, I_*) \, \tilde{\mathcal{B}}_{ij} \, \md v_* \, \md I_* & \geq 	\int_{B_\delta^c}  g(v_*, I_*) \, \tilde{\mathcal{B}}_{ij} \, \md v_* \, \md I_*  \geq \delta^{\g}    \int_{B_\delta^c}  g(v_*, I_*)  \, \md v_* \, \md I_* 
 	\\
% 	=  \delta^{\g}  \left( \int_{\bRfp}  g(v_*, I_*)    \md v_* \, \md I_*  -   \int_{B_\rho}  g(v_*, I_*)    \md v_* \, \md I_* \right)
 	&	=  \delta^{\g}  \left( \|g\|_{L^1_0}   -   \int_{B_\delta}  g(v_*, I_*)    \md v_* \, \md I_* \right).
 		\end{split}
 \end{equation}
 The next goal is to estimate the last integral in the previous inequality \eqref{pomocna 3}. To this end, we decompose the space for some suitable $K>1$ to be chosen later,
 \begin{align}
 	&  \int_{B_\delta}  g(v_*, I_*)    \md v_* \, \md I_*  = \int_{\bRfp}     \mathds{1}_{B_\delta(v,I)}  \left( \mathds{1}_{g(v_*, I_*) \geq K} + \mathds{1}_{g(v_*, I_*) \leq K}  \right) g(v_*, I_*) \,\md v_* \, \md I_*  \nonumber
 	  \\
 &	  \leq \int_{\bRfp}      \left( \tfrac{\log g(v_*, I_*)}{\log K} \mathds{1}_{g(v_*, I_*) \geq K} +  \mathds{1}_{B_\delta(v,I)} \mathds{1}_{g(v_*, I_*) \leq K}  \right) g(v_*, I_*) \,\md v_* \, \md I_*  \nonumber
 	  \\
 &	   \leq \int_{\bRfp}      \left(g(v_*, I_*) \tfrac{\left| \log g(v_*, I_*) \right|}{\log K}   +  \mathds{1}_{B_\delta(v,I)}  K \right)  \,\md v_* \, \md I_*  = \frac{H}{\log K} + K  \int_{\bRfp}       \mathds{1}_{B_\delta(v,I)}    \,\md v_* \, \md I_*. \label{pomocna 4}
 \end{align}
 The last integral can be explicitly computed,
 \begin{equation*}
 \int_{\bRfp}       \mathds{1}_{B_\delta(v,I)}    \,\md v_* \, \md I_* = \frac{16 \pi}{15} \frac{\mP^2}{\mu_{ij}}   \delta^5. 
 \end{equation*}
 Bringing this result back to \eqref{pomocna 4}, 
  \begin{equation}\label{pomocna 5}
  \int_{B_\delta}  g(v_*, I_*)    \md v_* \, \md I_*  \leq \frac{H}{\log K} + K  \frac{16 \pi}{15} \frac{\mP^2}{\mu_{ij}}   \delta^5 \leq \frac{ \|g\|_{L^1_0} }{2},
 \end{equation}
for the choice of $K>1$ such that
 \begin{equation*}
  \frac{H}{\log K} \leq \frac{ \|g\|_{L^1_0} }{4} \quad \Rightarrow \quad K \geq e^{{4 H}/{ \|g\|_{L^1_0}}},
 \end{equation*}
 and $\delta$ such that
 \begin{equation}\label{choice delta}
K  \frac{16 \pi}{15} \frac{\mP^2}{\mu_{ij}}   \delta^5 \leq  \frac{ \|g\|_{L^1_0} }{4} \quad \Rightarrow \quad \delta^5 \leq  \frac{15}{64 \pi} \frac{\mu_{ij}} {\mP^2}  \|g\|_{L^1_0}  \ e^{-{4 H}/{ \|g\|_{L^1_0}}}.
 \end{equation}
Thus, plugging \eqref{pomocna 5}  into \eqref{pomocna 3}, the final estimate in this step is obtained
 \begin{equation}	\label{lbl key 2}
		\int_{\bRfp}  g(v_*, I_*) \, \tilde{\mathcal{B}}_{ij} \, \md v_* \, \md I_* \geq  \delta^{\g}  \frac{ \|g\|_{L^1_0} }{2} =: L_2,
\end{equation}
with $\delta$ from \eqref{choice delta}.
 \subsection*{Step 3}  We gather estimates 	\eqref{lbl key 1} and 	\eqref{lbl key 2} from the previous steps, 
 \begin{equation}
 	\int_{\bRfp}  g(v_*, I_*) \, \tilde{\mathcal{B}}_{ij} \, \md v_* \, \md I_* \geq \max\left\{L_2,  L_1 \, \|g\|_{L^1_0} 	 \left( \frac{m_i}{2 \mP} |v|^2  + \frac{I}{\mP} \right)^{\!\!\frac{\g}{2}} - \|g\|_{L^1_{j, \g}} \right\}.
 \end{equation}
Using the pointwise estimate, valid for any nonnegative $a,b,c,x$
\begin{equation*}
\max\{-a+b\,x,c\}\geq \frac{c}{2}\big(1+\tfrac{b}{a+c}\,x\big)\geq\tfrac{c}{2}\min\{1,\tfrac{b}{a+c}\}\big(1+x\big)\,,
\end{equation*}
with $a=\|g\|_{L^1_{j, \g}}$, $b= L_1 \, \|g\|_{L^1_0}$, $c=L_2$, and $x=\big( \frac{m_i}{2 \mP} |v|^2  + \frac{I}{\mP} \big)^{\!\!\frac{\g}{2}}$, it follows that
\begin{equation*}
	\int_{\bRfp}  g(v_*, I_*) \, \tilde{\mathcal{B}}_{ij} \, \md v_* \, \md I_* \geq \tilde{c}^{lb}_{ij}[g] \left(  1+	 \left( \frac{m_i}{2 \mP} |v|^2  + \frac{I}{\mP} \right)^{\!\!\frac{\g}{2}} \right) \geq  \tilde{c}^{lb}_{ij}[g] \, \la v, I \ra_{i}^{\g}
\end{equation*}
 where the constant is
 \begin{equation}\label{c lower bound}
  \tilde{c}^{lb}_{ij}[g]=  \frac{L_2}{2} \min\left\{ 1, \frac{L_1 \, \|g\|_{L^1_0} }{L_2  +  \|g\|_{L^1_{j, \g}} } \right\}.
 \end{equation}
\end{proof}

As an immediate consequence, the following proposition holds.
\begin{proposition}\label{Prop: coll freq}
For $g$ satisfying assumptions of the previous Lemma \ref{Lemma Low b} and for the collision kernel $ \mathcal{B}_{ij}$ satisfying assumptions \eqref{p-p ass B},  the following lower bound on the collision frequency  $	\nu_{ij}[g](v,I) $ given in \eqref{coll fr}  holds
\begin{equation}\label{lbl coll fr}
	\nu_{ij}[g](v,I) \geq  \|b_{ij}\|_{L^1}  \, c^{lb}_{ij}[g] \,  \la v, I \ra_{i}^{\g},
\end{equation}
with the constant
\begin{equation*}
	c^{lb}_{ij}[g]  =  \tilde{c}^{lb}_{ij}[g]\ 	\|\tilde{b}_{ij}^{lb}\|_{L^1(d_{ij} \, \md r \, \md R)}.
\end{equation*}
\end{proposition}

\begin{proof} The proof follows by using assumption on $ \mathcal{B}_{ij}$ and   applying Lemma \ref{Lemma Low b},
	\begin{equation*}
		\begin{split}
	\nu_{ij}[g](v,I) &= \int_{\bRfp} \int_{\bS \times [0,1]^2} g(v_*, I_*) \,  \mathcal{B}_{ij} \, \dpo  \, \md \sigma \, \md r \, \md R \, \md v_* \, \md I_*
	\\ &\geq \int_{\bRfp} \int_{\bS \times [0,1]^2} g(v_*, I_*) \,  	b_{ij}(\hat{u}\cdot \sigma) \,  \tilde{b}_{ij}^{lb}(r, R) \,  \tilde{\mathcal{B}}_{ij}\,\dpo  \, \md \sigma \, \md r \, \md R \, \md v_* \, \md I_*
\\	& \geq  \|b_{ij}\|_{L^1}  \, c^{lb}_{ij}[g] \,  \la v, I \ra_{i}^{\g}.
		\end{split}
	\end{equation*}
\end{proof}

\section{Estimates on the collision   operator bi-linear forms}\label{Sec: coll op estimates}

\subsection{Estimates on bi-linear forms for the  gain part of the collision   operator }

In this section estimates from propositions \ref{Prop: Lp weak form} and \ref{Prop: Lp weak form bdd} are used to show a suitable control of a bi-linear form 
\begin{equation}\label{bi-lin form}
	\mathcal{Q}^+_{ij}[f, g] =
	\int_{\bRfp} \left(   Q^+_{ij}(f, g) \ f^{p-1} \ \la v,I \ra_i^{k \, p}  +   Q^+_{ji}(g, f) \ g^{p-1} \ \la v,I \ra_j^{k \, p} \right) \ \md v \, \md I,
\end{equation}
with the collision kernel \eqref{p-p ass B} having the angular part $	b_{ij}  \in L^1$.  Consequently, such an angular collision kernel can be decomposed without the loss of generality  in two suitable $L^{1}-L^{\infty}$ pieces.  For any $\eps>0$ we have the $\eps$-dependent decomposition:
\begin{equation}\label{bij L1 decomp}
	b_{ij} = b_{ij}^1 + b_{ij}^\infty, \quad \text{where}  \ b_{ij}^1  \in L^1 \ \text{satisfies} \ \| b_{ij}^1 \|_{\Ls} \leq \eps,  \ \text{and}  \  b_{ij}^\infty\in L^\infty,
\end{equation}
with the norm $\| b_{ij}^\infty \|_{\Linf}$ depending on $\eps$.

\begin{proposition}\label{Prop bi-linear form +}
	Let $f \in L^p_{i, {\g}/{p}+k} $ and $g \in L^p_{j, {\g}/{p}+k}$, for $k\geq0$, $\g\in [0,2]$, and $p \in [1,\infty)$, such that for the conjugate index $q$, the constant $	\cS(q)$ from \eqref{cond sing r} is finite.  In addition, let $H[f]$ and $H[g]$ as defined in \eqref{entropy H0} be finite.  The bi-linear form \eqref{bi-lin form} with the collision kernel satisfying assumption \eqref{p-p ass B} and having the  integrable angular part decomposed as in    \eqref{bij L1 decomp}, can be estimated, for any $\eps, \tilde{\eps} >0$ and $K > 1$,
\begin{multline}\label{Q+ ij bi-lin est}
	\mathcal{Q}^+_{ij}[f, g] \leq 
	4\pi \| b_{ij}^\infty \|_{\Linf}   \Bij
	\\
	+ \left( \eps \, \chfg +	4\pi \| b_{ij}^\infty \|_{\Linf}  \left( \frac{2\cLp(q)}{q} \, \tilde{\eps} + \frac{  \Cijfg }{(\log K)^{1/( k + \g +1)}}   \right) \right) \| f    \|^p_{L^p_{i, {\g}/{p}+k}}
	\\
	+ \left(  \eps \, \chgf + 	4\pi \| b_{ij}^\infty \|_{\Linf}  	\left(  \frac{2 c_{ji}(q)}{q} \, \tilde{\eps} + \frac{\Cjigf }{(\log K)^{1/( k + \g +1)}} \right)  \right) \,  \| g    \|_{L^p_{j, {\g}/{p}+k}}^p,
\end{multline}
where constants $\chfg$, $\Cijfg$ and $ \Bij$, that  are explicitly computed  along  the proof, depend on $H[f]$, $H[g]$ and the $L^1_{ k+\g +1}$-norms related to both $f$ and $g$. 
\end{proposition}

\begin{proof} The proof is divided into three steps.
\subsection*{Step 1} First consider the bi-linear form    \eqref{bi-lin form} taking $b_{ij}^1$ whose norm is $ \| b_{ij}^1 \|_{\Ls} \leq \eps$. Applying Proposition \ref{Prop: Lp weak form}, one gets
\begin{multline}\label{pomocna 8}
	\mathcal{Q}^+_{ij}[f, g] 
	\leq 2^{\frac{p-1}{2p}}  \,\eps \left( \Big(   \| f    \|_{L^p_{i, {\g}/{p}+k}}  \|  	g    \|_{L^1_{j, k+\g}} 
	+    \| 	f  \|_{L^1_{i, k+\g}}  \| g   \|_{L^p_{j, {\g}/{p}+k}}    \Big) \cLp \| f  \|_{L^p_{i,  {\g}/{p}+k}}^{p-1} 
\right.	\\ \left.
	+  \Big(   \| g   \|_{L^p_{j, {\g}/{p}+k}}  \|  	f    \|_{L^1_{i, k+\g}} 
	+    \| 	g  \|_{L^1_{j, k+\g}}  \| f   \|_{L^p_{i, {\g}/{p}+k}}    \Big) c_{ji} \| g  \|_{L^p_{j,  {\g}/{p}+k}}^{p-1} \right),
\end{multline}
where the left-hand side is to be understood as a form  \eqref{bi-lin form}  for $b_{ij}^1$. For the mixed terms, the Young's inequality yields
\begin{equation}\label{Y f g}
\begin{split}
  \| f    \|^{p-1}_{L^p_{i, {\g}/{p}+k}}  \| g    \|_{L^p_{j, {\g}/{p}+k}}
\leq   \tfrac{1}{q} \| f    \|^{p}_{L^p_{i, {\g}/{p}+k}}  + \tfrac{1}{p} \| g    \|_{L^p_{j, {\g}/{p}+k}}^p,
\end{split}
\end{equation}
and therefore the bilinear form \eqref{pomocna 8}  for the integrable part $b_{ij}^1$ of the angular kernel  becomes
\begin{multline}\label{Q+ ij L1 est}
	\mathcal{Q}^+_{ij}[f, g] 
%	\leq 2^{\frac{p-1}{2p}}  \,\eps \left\{ \Big(   \| f    \|^p_{L^p_{i, {\g}/{p}+k}}  \|  	g    \|_{L^1_{j, k+\g}} 
%	+  \| 	f  \|_{L^1_{i, k+\g}}  \left(   \tfrac{1}{q} \| f    \|^{p}_{L^p_{i, {\g}/{p}+k}}  + \tfrac{1}{p} \| g    \|_{L^p_{j, {\g}/{p}+k}}^p \right) \Big) \cLp 
%	\right.	\\ \left.
%	+  \Big(   \| g   \|^p_{L^p_{j, {\g}/{p}+k}}  \|  	f    \|_{L^1_{i, k+\g}} 
%	+    \| 	g  \|_{L^1_{j, k+\g}}  \left(   \tfrac{1}{q} \| g    \|^{p}_{L^p_{j, {\g}/{p}+k}}  + \tfrac{1}{p} \| f    \|_{L^p_{i, {\g}/{p}+k}}^p \right)      \Big) c_{ji} \right\}
		\leq 2^{\frac{p-1}{2p}}  \,\eps \left\{  \| f    \|^p_{L^p_{i, {\g}/{p}+k}}  \Big( \cLp \|  	g    \|_{L^1_{j, k+\g}} + \tfrac{\cLp}{q}   \| 	f  \|_{L^1_{i, k+\g}} + \tfrac{c_{ji}}{p}  \| 	g  \|_{L^1_{j, k+\g}}  \Big)
	\right.	\\ \left.
	+  \| g   \|^p_{L^p_{j, {\g}/{p}+k}} \Big( c_{ji}  \|  	f    \|_{L^1_{i, k+\g}} 
	+  \tfrac{c_{ji}}{q}     \| 	g  \|_{L^1_{j, k+\g}}   +   \tfrac{\cLp}{p}  \| 	f  \|_{L^1_{i, k+\g}}    \Big)    \right\}
	\\
	=: \eps \left( \chfg  \,  \| f    \|^p_{L^p_{i, {\g}/{p}+k}} +  \chgf \,  \| g   \|^p_{L^p_{j, {\g}/{p}+k}}  \right),
\end{multline}
where the constant is abbreviated as 
\begin{equation}
	\chfg = 2^{\frac{p-1}{2p}} \left( \cLp(q) \|  	g    \|_{L^1_{j, k+\g}} + \tfrac{\cLp(q)}{q}   \| 	f  \|_{L^1_{i, k+\g}} + \tfrac{c_{ji}(q)}{p}  \| 	g  \|_{L^1_{j, k+\g}} \right).
\end{equation}
\subsection*{Step 2}  Next consider the form  \eqref{bi-lin form}  for the part $b_{ij}^\infty \in L^\infty$. We study separately two terms of \eqref{bi-lin form} and moreover for each term the cases $\hat{u}\cdot\sigma \geq 0 $ and $\hat{u}\cdot\sigma \leq 0 $ are dissociated, or in other words, decomposition \eqref{bij decom} is relevant,
 \begin{equation}\label{bij inf decom}
	b_{ij}^\infty   = b_{ij}^\infty  \, \mathds{1}_{\hat{u}\cdot \sigma \geq 0}  + b_{ij}^\infty  \, \mathds{1}_{\hat{u}\cdot \sigma \leq 0} =:  (b_{ij}^\infty )^+   + (b_{ij}^\infty )^-.
\end{equation}
Thus,   the first term in \eqref{bi-lin form} , for some $K>1$ and $\ell \geq0$ to be determined later, is decomposed on four terms as follows
\begin{multline}\label{pomocna 9}
	\int_{\bRfp} Q_{ij}^+(f, g) \,f^{p-1} \ \la v,I \ra_i^{k \, p} \, \md v\, \md I  
	\\
	=	\int_{\bRfp} \left( Q_{ij}^+(f, g \mathds{1}_{g \la \cdot \ra_j^\ell \geq K})   +  Q_{ij}^+(f, g \mathds{1}_{g \la \cdot \ra_j^\ell \leq K}) \right)  \mathds{1}_{\hat{u} \cdot \sigma \geq 0}  \,f^{p-1} \ \la v,I \ra_i^{k \, p} \, \md v\, \md I
	\\
	+ 	\int_{\bRfp} \left( Q_{ij}^+(f  \mathds{1}_{f \la \cdot \ra_i^\ell \geq K}, g)   +  Q_{ij}^+(f\mathds{1}_{f \la \cdot \ra_i^\ell \leq K}, g) \right)  \mathds{1}_{\hat{u} \cdot \sigma \leq 0}  \,f^{p-1} \ \la v,I \ra_i^{k \, p} \, \md v\, \md I
\\ =: T_1 + T_2 +T_3 +T_4.
\end{multline}
The goal is to apply Proposition \ref{Prop: Lp weak form bdd} and its all four estimates. In either case, for $\chi=f^{p-1}$,
\begin{equation*}
		\| f^{p-1}  \|_{L^{\q}_{i, (k p+ {\g})/{\q}}}  =  \| f    \|^{p-1}_{L^p_{i, {\g}/{p}+k}}.
\end{equation*}
For the  term $T_1$,   \eqref{Lp weak form inf +}$_1$ imply
	\begin{equation}\label{pomocna 10}
%	\int_{\bRfp}  Q_{ij}^+(f, g \mathds{1}_{g \la \cdot \ra_j^\ell \geq K})   \,f^{p-1} \ \la v,I \ra_i^{k \, p} \, {\color{blue} \mathds{1}_{\hat{u} \cdot \sigma \geq 0}} \, \md v\, \md I   
%	\\
T_1	\leq  4\pi \| (b_{ij}^{ \infty})^+ \|_{\Linf}  \,  \cLp\,    \| f    \|^p_{L^p_{i, {\g}/{p}+k}}  \|  	g \mathds{1}_{g \la \cdot \ra_j^\ell \geq K}   \|_{L^1_{j, k+\g}}.
\end{equation}
Note that taking logarithm of $g \la \cdot \ra_j^\ell \geq K$ implies $1\leq  \frac{\log(g \la \cdot \ra_j^\ell) }{\log K}$, and thus H\"older inequality for the pair   $(k + \g +1, \tfrac{k + \g +1}{k+\g})$ yields
\begin{align}\label{est entropy}
		\|  	g \mathds{1}_{g \la \cdot \ra_j^\ell \geq K}   \|_{L^1_{j, k+\g}} 
		& \leq 	\|  	g \, \la \cdot \ra_j^{k+\g}  \Bigg(\frac{\log(g \la \cdot \ra_j^\ell) }{\log K}\Bigg)^{\frac{1}{k + \g +1}} \mathds{1}_{g \la \cdot \ra_j^\ell \geq K}   \|_{L^1}  \nonumber
		%\leq \frac{1}{(\log K)^{1/r}}	\| 	g \la \cdot \ra_j^{(k+\g) \, r'}  \|_{L^1}^{1/r'}  	\| g	\, \log(g \la \cdot \ra_j^\ell)    \|_{L^1}^{1/r} 
		 \\
		& \leq \frac{1}{(\log K)^{1/( k + \g +1)}}	\| 	g \la \cdot \ra_j^{k+\g+1 }  \|_{L^1}^{\frac{ k + \g}{ k + \g +1}}  	\| g	\, \log(g \la \cdot \ra_j^\ell)    \|_{L^1}^{\frac{1}{k + \g +1}}.
\end{align}
Next, we invoke the entropy by estimating 
\begin{equation*}
g	\, \left| \log(g \la \cdot \ra_j^\ell) \right|  \leq g \left| \log g \right| + \frac{\ell}{2} \log \la \cdot \ra_j^2 \leq  g \, \left| \log g \right| + \frac{\ell}{2}    \la \cdot \ra_j^2,
\end{equation*}
which for \eqref{est entropy} implies
\begin{equation}\label{pomocna 12}
	\|  	g \mathds{1}_{g \la \cdot \ra_j^\ell \geq K}   \|_{L^1_{j, k+\g}} 
	\leq \frac{1}{(\log K)^{1/( k + \g +1)}} \cgj
\end{equation}
where the constant $\cgj$ depends only on $L^1_j$ norms of $g$ and entropy $H$ related to $g$ as defined in \eqref{entropy H0}, more precisely,
\begin{equation}\label{cgj}
\cgj = 	\| 	g    \|_{L^1_{j, k+\g+1}}^{\frac{ k + \g}{ k + \g +1}}  \left( H[g] + \frac{\ell}{2}	\| 	g    \|_{L^1_{j, 2}} \right)^{\frac{1}{k + \g +1}},
\end{equation}
where $\ell$ will be chosen later.
Thus, \eqref{pomocna 10} becomes
	\begin{equation}\label{T1 est}
T_1	\leq \frac{\cgj}{(\log K)^{1/( k + \g +1)}}  4\pi \| (b_{ij}^\infty)^+ \|_{\Linf}  \,  \cLp \,  \| f    \|^p_{L^p_{i, {\g}/{p}+k}}.
\end{equation}
We proceed similarly for the term $T_3$. Namely, applying    \eqref{Lp weak form inf -}$_1$ and using the  arguments of \eqref{pomocna 12},
  	\begin{equation}\label{pomocna 11}
  		\begin{split}
  	T_3 &
  	\leq  4\pi \| ( b_{ij}^{ \infty} )^-  \|_{\Linf}  \,  \cLp\,    \| f    \|^{p-1}_{L^p_{i, {\g}/{p}+k}}  \| g    \|_{L^p_{j, {\g}/{p}+k}}  \|  	f \mathds{1}_{f \la \cdot \ra_i^\ell \geq K}   \|_{L^1_{i, k+\g}}
  	\\ 	
  	&	\leq  \frac{\cfi}{(\log K)^{1/( k + \g +1)}}   4\pi \| ( b_{ij}^{ \infty} )^-  \|_{\Linf}  \,  \cLp\,    \| f    \|^{p-1}_{L^p_{i, {\g}/{p}+k}}  \| g    \|_{L^p_{j, {\g}/{p}+k}}.
  	\end{split}
  \end{equation}
Then, after Young's inequality, $T_3$ becomes 
 	\begin{equation}\label{T3 est}
	\begin{split}
		T_3 
	\leq  \frac{\cfi}{(\log K)^{1/( k + \g +1)}}   4\pi \| ( b_{ij}^{ \infty} )^-  \|_{\Linf}  \,  \cLp\,  \left( \tfrac{1}{q} \| f    \|^{p}_{L^p_{i, {\g}/{p}+k}}  + \tfrac{1}{p} \| g    \|_{L^p_{j, {\g}/{p}+k}}^p \right).
	\end{split}
\end{equation}
For term $T_2$,  \eqref{Lp weak form inf +}$_2$ yields
	\begin{multline*}
%	\int_{\bRfp}  Q_{ij}^+(f, g \mathds{1}_{g \la \cdot \ra_j^\ell \leq K})   \,f^{p-1} \ \la v,I \ra_i^{k \, p} \, \md v\, \md I   
%	\\
T_2	\leq  4\pi \| (b_{ij}^\infty)^+ \|_{\Linf}  \,  \cLp\,    \|  g \mathds{1}_{g \la \cdot \ra_j^\ell \leq K}    \|_{L^p_{j, k+\g}}  \|  	f    \|_{L^1_{i, {\g}/{p}+k}} \, 	\| f  \|_{L^{p}_{i,  {\g}/{p}+k}}^{p-1}  
	\\
	\leq  4\pi \| (b_{ij}^\infty)^+ \|_{\Linf}  \,  \cLp\, K \,  \|   \la \cdot \ra_j^{k+\g-\ell}  \|_{L^p }    \|  	f    \|_{L^1_{i, {\g}/{p}+k}} \,	\| f  \|_{L^{p}_{i,  {\g}/{p}+k}}^{p-1}, 
\end{multline*}
for some  $\ell$ satisfying
\begin{equation}\label{cond ell}
	(   k + \g-\ell) p <-5, \quad \text{for instance } \ \ell = k + \gamma_{ij} + 6,
\end{equation}
for which 
\begin{equation}
 \|   \la \cdot \ra_j^{k+\g-\ell}  \|_{L^p }^p   =  \|   \la \cdot \ra_j^{-6}  \|_{L^p }^p  \leq    \|   \la \cdot \ra_j^{-6}  \|_{L^1 }   = \frac{\pi^2}{2}  \frac{\mP^{5/2}}{m_j^{3/2} }  =: \cchj.
\end{equation}
Therefore, Young's inequality implies, for some $\tilde{\eps}>0$,
\begin{equation}\label{T2 est}
T_2	\leq 4\pi \| (b_{ij}^\infty)^+ \|_{\Linf}  \cLp \left( \frac{ K^p}{p \, \tilde{\eps}^{p-1}}   \,\cchj   \|  	f    \|^p_{L^1_{i, {\g}/{p}+k}} + \frac{\tilde{\eps}}{q} 	\| f  \|_{L^{p}_{i,  {\g}/{p}+k}}^{p} \right).
\end{equation}
Similarly, for $T_4$,   \eqref{Lp weak form inf -}$_2$  implies, for $\ell\geq 0$ from \eqref{cond ell},
	\begin{align}
	T_4	&\leq  4\pi \| ( b_{ij}^{ \infty} )^- \|_{\Linf}  \,  \cLp\,    \|  f \mathds{1}_{f \la \cdot \ra_i^\ell \leq K}    \|_{L^p_{i, k+\g}}  \|  	g    \|_{L^1_{j, {\g}/{p}+k}} \, 	\| f  \|_{L^{p}_{i,  {\g}/{p}+k}}^{p-1}  \nonumber
	\\
&	\leq  4\pi \|  ( b_{ij}^{ \infty} )^-\|_{\Linf}  \,  \cLp\, K \,  \|   \la \cdot \ra_i^{k+\g-\ell}  \|_{L^p }    \|  	g    \|_{L^1_{j, {\g}/{p}+k}} \,	\| f  \|_{L^{p}_{i,  {\g}/{p}+k}}^{p-1} \nonumber
	\\&	\leq  4\pi \|  ( b_{ij}^{ \infty} )^-\|_{\Linf} \cLp \left(   \frac{K^p }{p \, \tilde{\eps}^{p-1}}      \cchi  \|  	g    \|^p_{L^1_{j, {\g}/{p}+k}}   + \frac{\tilde{\eps}}{q} 	\| f  \|_{L^{p}_{i,  {\g}/{p}+k}}^{p} \right)\,, \label{T4 est}
\end{align}
where for the last estimate Young's inequality is applied.

Gathering \eqref{T1 est}, \eqref{T3 est}, \eqref{T2 est} and \eqref{T4 est},
\begin{align*}
&	\int_{\bRfp} Q_{ij}^+(f, g) \,f^{p-1} \ \la v,I \ra_i^{k \, p} \, \md v\, \md I  
	  \leq \left\{ 4 \pi \, \cLp \Big(  \| (b_{ij}^\infty)^+ \|_{\Linf}  +  \| (b_{ij}^\infty)^- \|_{\Linf}  \Big) \frac{\tilde{\eps}}{q} 
	\right. \\ & \left.  
	\qquad \qquad
	  + \frac{4\pi  \,  \cLp}{(\log K)^{1/( k + \g +1)}}   \left( \| (b_{ij}^\infty)^+ \|_{\Linf} \,  \cgj +  \tfrac{1}{q} \, \| (b_{ij}^\infty)^- \|_{\Linf} \,  \cfi \right) \right\}\| f    \|^p_{L^p_{i, {\g}/{p}+k}}
	  \\
&\qquad \qquad	  +	 \frac{ 4\pi \, \cLp }{(\log K)^{1/( k + \g +1)}} \tfrac{1}{p} \, \|  (b_{ij}^{ \infty})^-  \|_{\Linf}  \,  \cfi \, \| g    \|_{L^p_{j, {\g}/{p}+k}}^p
	  \\
&	  + \frac{4\pi  \,  \cLp \, K^p}{p \, \tilde{\eps}^{p-1}} \left( \| (b_{ij}^\infty)^+ \|_{\Linf} \,  \cchj     \,  \|  	f    \|^p_{L^1_{i, {\g}/{p}+k}}  + \| (b_{ij}^\infty)^- \|_{\Linf}  \, \cchi \,   \|  	g    \|_{L^1_{j, {\g}/{p}+k}}^p \right).
\end{align*}
The last estimate makes clear that back scattering manifested here through $  \| (b_{ij}^\infty)^- \|_{\Linf} $ brings into play $L^p$ norm of $g$, making essential to consider the bilinear form \eqref{bi-lin form}. %or putting on evidence the strong coupling 
However, both $L^\infty$ norms of $(b_{ij}^\infty)^+$ and $(b_{ij}^\infty)^- $ can be bounded, making the estimate simpler, 
	  \begin{align*}
	  	&	\int_{\bRfp} Q_{ij}^+(f, g) \,f^{p-1} \ \la v,I \ra_i^{k \, p} \, \md v\, \md I  
\\	  &	\leq 
 4\pi \| b_{ij}^\infty \|_{\Linf} \, \cLp(q) \left\{    \frac{ K^p}{p \, \tilde{\eps}^{p-1}} \left(  \cchj    \|  	f    \|^p_{L^1_{i, {\g}/{p}+k}}  + \cchi   \|  	g    \|_{L^1_{j, {\g}/{p}+k}}^p \right).  
\right. \\ & \left.
\qquad \qquad \qquad \qquad
+ \left( 2 \frac{\tilde{\eps}}{q} + \frac{       \cgj +  \tfrac{1}{q} \cfi   }{(\log K)^{1/( k + \g +1)}}   \right) \| f    \|^p_{L^p_{i, {\g}/{p}+k}}
	  +  \frac{   \tfrac{1}{p} \cfi }{(\log K)^{1/( k + \g +1)}}    \,  \| g    \|_{L^p_{j, {\g}/{p}+k}}^p \right\}.
\end{align*}

The second term in \eqref{bi-lin form} follows the same steps as the first one, and therefore, the bi-linear form \eqref{bi-lin form} for $b_{ij}^\infty$ becomes
\begin{multline}\label{Q+ ij Linf est}
	\mathcal{Q}^+_{ij}[f, g] \leq 
   4\pi \| b_{ij}^\infty \|_{\Linf} \left\{  \Bij
 	\right. \\  \left.+ \left(  \frac{2\cLp(q)}{q} \, \tilde{\eps}+ \frac{  \Cijfg   }{(\log K)^{1/( k + \g +1)}}   \right) \| f    \|^p_{L^p_{i, {\g}/{p}+k}}
	 	+	\left( \frac{2 c_{ji}(q)}{q} \, \tilde{\eps} + \frac{  \Cjigf }{(\log K)^{1/( k + \g +1)}}   \right) \,  \| g    \|_{L^p_{j, {\g}/{p}+k}}^p
 \right\}\,,
\end{multline}
where the involved constants are 
\begin{equation}\label{constants C2 and B}
	\begin{split}
\Cijfg &= \cLp(q)  \, \cgj +    \tfrac{\cLp(q)}{q} \, \cfi  +  \tfrac{c_{ji}(q) }{p} \, \cgj, \\
\Bij  &= 	 \frac{ K^p}{ p \, \tilde{\eps}^{p-1}} \left( \cLp(q)  + c_{ji}(q)  \right)  \left(    \cchj    \|  	f    \|^p_{L^1_{i, {\g}/{p}+k}}  + \cchi   \|  	g    \|_{L^1_{j, {\g}/{p}+k}}^p\right)   
\end{split}
\end{equation}
with notation \eqref{cgj}.

\subsection*{Step 3}  Finally, gathering estimates for bi-linear forms taking integrable \eqref{Q+ ij L1 est} and bounded \eqref{Q+ ij Linf est} angular kernels,   the stated estimate \eqref{Q+ ij bi-lin est} for the bi-linear form \eqref{bi-lin form} with the angular kernel satisfying \eqref{bij L1 decomp} is obtained, which concludes the proof.
\end{proof}

\subsection{Estimates on the collision operator written in a bi-linear form}
Defining the bi-linear form of the loss operator, 
 \begin{equation}\label{bi-lin form -}
 		\mathcal{Q}^-_{ij}[f, g] 
=	\int_{\bRfp} \left(   Q^-_{ij}(f, g) \ f^{p-1} \ \la v,I \ra_i^{k \, p}  +   Q^-_{ji}(g, f) \ g^{p-1} \ \la v,I \ra_j^{k \, p} \right) \ \md v \, \md I,
 \end{equation}
the bi-linear form of the complete collision operator  \eqref{p-p coll operator} is written as
\begin{equation}\label{bi-linear form total}
	\mathcal{Q}_{ij}[f, g]  = 	\mathcal{Q}^+_{ij}[f, g]  - 	\mathcal{Q}^-_{ij}[f, g].
\end{equation}
 This bi-linear form is estimated as follows.
 
 \begin{proposition}\label{Prop bi-linear form} 
 	Assume $f \in L^p_{i, {\g}/{p}+k}$ and $g \in L^p_{j, {\g}/{p}+k}$, for $k\geq0$, $\g \in [0,2]$, and $p \in [1,\infty)$, such that for the conjugate index $q$ the constant $\cS(q)$ from \eqref{cond sing r} is finite.  In addition, assume $H[f]$ and $H[g]$ as defined in \eqref{entropy H0} to be finite.  The bi-linear form \eqref{bi-linear form total} with the collision kernel \eqref{p-p ass B} and integrable angular part $b_{ij}$ can be estimated as
\begin{equation}\label{bi-linear final}
	\mathcal{Q}_{ij}[f, g]  \leq 
	-   \frac{1}{2}  \|b_{ij}\|_{L^1} \left(  \ c^{lb}_{ij}[g] \  \| f    \|^p_{L^p_{i, {\g}/{p}+k}}
	+ \ c^{lb}_{ji}[f] \  \| g    \|_{L^p_{j, {\g}/{p}+k}}^p \right) +   4\pi \| b_{ij}^\infty \|_{\Linf} \, \Bij\,,
\end{equation}
where constants are from Lemma \ref{Lemma Low b} and Proposition \ref{Prop bi-linear form +}.
 \end{proposition}
 \begin{proof}
Using the local form of the loss operator \eqref{coll loss operator}, the bi-linear form \eqref{bi-lin form -} can be estimated by applying Proposition \ref{Prop: coll freq},
 \begin{equation}
 	\begin{split}
 	\mathcal{Q}^-_{ij}[f, g] 
 	&=	\int_{\bRfp} \left(   \nu_{ij}[g](v,I) \ f^{p} \ \la v,I \ra_i^{k \, p}  +   \nu_{ji}[f](v,I) \ g^{p} \ \la v,I \ra_j^{k \, p} \right) \ \md v \, \md I
 	\\
 		& \geq   \|b_{ij}\|_{L^1}  \left(    c^{lb}_{ij}[g] \ \| f \|_{L^p_{i, \g/p +k}}^p   +   c^{lb}_{ji}[f] \ \| g \|_{L^p_{j, \g/p +k}}^p   \right).
 	\end{split}
 \end{equation}
This estimate together with \eqref{Q+ ij bi-lin est} yields the following bound on  the complete bi-linear form \eqref{bi-linear form total} 
\begin{multline}
 	\mathcal{Q}_{ij}[f, g]  \leq 4\pi \| b_{ij}^\infty \|_{\Linf} 	\Bij\\
 	- \left(   \|b_{ij}\|_{L^1}  c^{lb}_{ij}[g] -  \eps \, \chfg -	4\pi \| b_{ij}^\infty \|_{\Linf}  \left( \frac{2\cLp}{q} \, \tilde{\eps} +\frac{  \Cijfg }{(\log K)^{1/( k + \g +1)}}   \right) \right) \| f    \|^p_{L^p_{i, {\g}/{p}+k}}
 	\\
 	- \left(  \|b_{ij}\|_{L^1}  c^{lb}_{ji}[f]  -   \eps \, \chgf - 	4\pi \| b_{ij}^\infty \|_{\Linf}  	\left(\frac{2 c_{ji}}{q} \, \tilde{\eps}  + \frac{\Cjigf }{(\log K)^{1/( k + \g +1)}} \right)  \right) \,  \| g    \|_{L^p_{j, {\g}/{p}+k}}^p,
\end{multline}
The next step is the choice of constants $\eps$, $\tilde{\eps}$ and $K$ such that positive coefficients of $L^p$ norms coming from the gain operator  are absorbed into the negative ones coming from the loss operator. To that end, first choose $\eps$
\begin{equation}
	\eps = \frac{1}{4}  \|b_{ij}\|_{L^1} \min\left\{ \frac{  c^{lb}_{ij}[g] }{ \chfg},  \frac{   c^{lb}_{ji}[f] }{ \chgf} \right\}.
\end{equation}
Note that  such a choice of $\eps$ fixes the decomposition \eqref{bij L1 decomp} and in particular  the norm $ \| b_{ij}^\infty \|_{\Linf}$. Furthermore, we   proceed with a choice of $	\tilde{\eps}$ and $K$. Namely, taking
 \begin{equation}\label{K eps}
 	\begin{split}
	\tilde{\eps} &= \frac{q}{64 \, \pi} \frac{ \|b_{ij}\|_{L^1}  }{  \| b_{ij}^\infty \|_{\Linf}}   \min\Big\{ \frac{c^{lb}_{ij}[g]}{\cLp(q)},   \frac{c^{lb}_{ji}[f]}{ c_{ji}(q) }   \Big\},\\
 K &= \exp \left( 32 \pi  \frac{  \| b_{ij}^\infty \|_{\Linf}}{ \|b_{ij}\|_{L^1}  }   \max\left\{  \frac{\Cijfg}{c^{lb}_{ij}[f]},  \frac{\Cjigf}{c^{lb}_{ji}[g]}    \right\} \right)^{k + \g +1} 
 \end{split}
\end{equation}
yields the statement \eqref{bi-linear final}.
\end{proof}

\section{Propagation of $L^p$ norms of the solution to the Boltzmann system}\label{Sec: Prop}

The previous Proposition \ref{Prop bi-linear form} that states the bound on the  bi-linear form of the collision operator, allows to derive the estimate on the vector valued collision operator. This estimate provides an ODI for the solution of the Boltzmann system \eqref{BE vector} enabling to conclude an integrability propagation result.\\

Before entering in the details, we note that the inequality $x|\log(x)|\leq x^{\frac34} + c_p\, x^{p}$, valid for any $x\geq0$, $p\in(1,\infty)$, implies that $\F_0\in L^{1}_{2}\cap L^{p}$ guarantees that $\mathcal{H}(\F_0)<\infty$.  For the particular case $p=\infty$, if $0\leq x \leq x_0$ then $$x|\log(x)|\leq \max_{0\leq y \leq x_0}(y^\frac14 |\log(y)|) \, x^{\frac34}\,,$$ which implies again that $\F_0\in L^{1}_{2}\cap L^{\infty}$ guarantees that $\mathcal{H}(\F_0)<\infty$. Then, for the solutions of the system estimate \eqref{H finite} implies that $\sup_{t\geq0}H[\F]$ is finite with the condition $\F_0\in L^{1}_{2}\cap L^{p}$, $p>1$.

\begin{theorem} Let $\F$ be a solution of the Boltzmann system of equations in the sense of \ref{Sec: solutions} with initial data 
	\begin{equation*}
 \| \F_0    \|_{L^p_{k}}< \infty  \qquad \text{and} \qquad \| \F_0    \|_{L^1_{k+ \gm +1}}< \infty,
	\end{equation*}
 for  $\gm$ from \eqref{gamma w}, $k \geq 0$ and  $p \in (1,\infty)$ such that the constant $\cS(q)$ related to the conjugate index $q$ given in \eqref{cond sing r} is finite for any pair of indices $i,j \in \left\{1, \dots, P\right\}$. Then,  
\begin{equation}\label{Lp propagation}
	\| \F    \|^p_{L^p_{k}}(t)  \leq \max\left\{  \| \F_0    \|^p_{L^p_{k}}, \frac{B}{A}  \right\}, \quad \text{for} \ t\geq 0,
\end{equation}
where the constants are
\begin{equation}\label{const A B}
  A =    \min_{1 \leq i \leq P }	\sum_{j=1}^{P}  
	\|b_{ij}\|_{L^1}  \ c^{lb}_{ij}[f_j] , \quad	B =   \sum_{i, j=1}^{P} 4\pi \| b_{ij}^\infty \|_{\Linf} 	 B_{ij}[f_i, f_j] ,
\end{equation}
with $c^{lb}_{ij}$ from Lemma \ref{Lemma Low b} and $B_{ij}$ from  Proposition \ref{Prop bi-linear form +}.\\

When $p=\infty$, the following result holds 
\begin{equation}\label{Lp propagation inf}
	\| \F    \|_{L^\infty_{k}}(t)  \leq \max\left\{  \| \F_0    \|_{L^\infty_{k}}, {\tilde{B}}  \right\}, \quad \text{for} \ t\geq 0,
\end{equation}
with the constant $\tilde{B}$ explicitly given in the proof (which requires $\cS(1)<\infty$).

\end{theorem}

\begin{proof}
Multiplying the Boltzmann equation \eqref{BE i} for $f_i$ with $f_i^{p-1} \ \la v,I \ra_i^{k \, p}$  and then integrating with respect to $(v,I)$ and summing over all mixture components $i=1,\dots,P$ gives an ODE for the $L^p$ polynomially weighted norm of the vector valued solution $\F$,  
\begin{equation}\label{ODE}
\frac{1}{p}  \| \F    \|^p_{L^p_{k}} = 	\sum_{i, j=1}^{P} \int_{\bRfp} Q_{ij}(f_i, f_j)(v,I) \ f_i^{p-1} \ \la v,I \ra_i^{k \, p}.
\end{equation}
The right-hand side can be related to the bi-linear form \eqref{bi-linear form total}. Note that $L^1$ theory for the solution of the Boltzmann system \eqref{BE vector} established in \cite{MPC-A-G} and, in particular, the propagation result stated in Proposition \ref{Prop L1},   implies that the solution $\F$ will have finite $L^1_{k+\gm +1}$ norm at any time $t>0$, since initially so, by the assumption of this theorem. In addition, the aforementioned considerations previous to this theorem imply the finiteness of $\sup_{t\geq0}H[\F]$. Thus, all constants from Proposition \ref{Prop bi-linear form} are finite, and the estimate \eqref{bi-linear final} on the bi-linear form implies 
\begin{multline}
2	\sum_{i, j=1}^{P} \int_{\bRfp} Q_{ij}(f_i, f_j)(v,I) \ f_i^{p-1} \ \la v,I \ra_i^{k \, p} =  	\sum_{i, j=1}^{P}   	\mathcal{Q}_{ij}[f_i, f_j] 
	\\
	 \leq 
	\sum_{i, j=1}^{P}   \left(
	- \frac{1}{2}  \|b_{ij}\|_{L^1}  \ c^{lb}_{ij}[f_j] \  \| f_i    \|^p_{L^p_{i, {\g}/{p}+k}}
	-  \frac{1}{2}  \|b_{ij}\|_{L^1} \ c^{lb}_{ji}[f_i] \  \| f_j    \|_{L^p_{j, {\g}/{p}+k}}^p  + 4\pi \| b_{ij}^\infty \|_{\Linf}  	B_{ij}[f_i, f_j] \right)
	\\
=
\sum_{i, j=1}^{P}   \left(
	-    \|b_{ij}\|_{L^1}  \ c^{lb}_{ij}[f_j] \  \| f_i    \|^p_{L^p_{i, {\g}/{p}+k}} + 4\pi \| b_{ij}^\infty \|_{\Linf} B_{ij}[f_i, f_j]  \right)
	\leq 
% \sum_{i, j=1}^{P}  C_3^{ij}
%	-
%  \left(	\min_{1 \leq i \leq P }	\sum_{j=1}^{P}  
%	 \|b_{ij}\|_{L^1}  \ c^{lb}_{ij}[f_j] \right)    \| \F    \|^p_{L^p_{k}} = 
	 - A  \| \F    \|^p_{L^p_{k}}  + B,
\end{multline}
where the last estimate is due to monotonicity of brackets \eqref{brackets} for $\g \geq 0$ and for constants \eqref{const A B}.
Therefore, \eqref{ODE} becomes an ODI,
\begin{equation}
	\partial_t  \| \F    \|^p_{L^p_{k}}  \leq - \frac{ A \, p}{2}  \| \F    \|^p_{L^p_{k}}  + \frac{B \, p}{2}.
\end{equation}
Thus,  direct integration implies the statement \eqref{Lp propagation}.\\

For the case $p=\infty$, we study dependence on $p$ of constants involved in inequality \eqref{Lp propagation} recalled here as
\begin{equation}\label{p root} 
\| \F    \|_{L^p_{k}}(t)  \leq \max\left\{  \| \F_0    \|_{L^p_{k}}, \left(\frac{B}{A}\right)^{1/p}  \right\}. 
\end{equation}
First, recalling the constant $B_{ij}[f_i,f_j]$ from \eqref{constants C2 and B} and since $1/p<1$, it holds that
\begin{equation*}
B^{1/p} \leq\sum_{i, j=1}^P	 \frac{ K}{  \tilde{\eps}^{1-1/p}} \left(\frac{ \cLp(q)  + c_{ji}(q)}{p}  \right)^{1/p}  \left(    \cchj^{1/p}    \|  	f_i    \|_{L^1_{i, {\g}/{p}+k}}  + \cchi^{1/p}   \|  	f_j    \|_{L^1_{j, {\g}/{p}+k}}\right).
\end{equation*}
Second, by hypothesis $\F_0\in L^{1}_{k}\cap L^\infty_{k}$, it follows that
\begin{equation*}
\lim_{p\rightarrow\infty} \| \F_0    \|_{L^p_{k}} = \| \F_0 \|_{ L^{\infty}_{k} }\,,
\end{equation*}
and third, since the constant $A$ does not depend on $p$, $\lim_{p\rightarrow\infty} A^{1/p}= 1$.  Therefore, sending $p\rightarrow\infty$ in \eqref{p root} one has
\begin{equation*}
\limsup_{p}\| \F    \|_{L^p_{k}}(t) \leq \max\left\{  \| \F_0    \|_{L^\infty_{k}}\, ,  \sum_{i, j=1}^P	 \frac{ 2 \, K}{  \tilde{\eps} }  \,     \|  	f_i    \|_{L^1_{i, k}} \right\}\,,
\end{equation*}
where $K$ and  $ \tilde{\eps} $ are from \eqref{K eps} with $q=1$ (note the importance of $\cS(1)<\infty$ as we take the limit).  This is enough to conclude that $\F(t) \in L^{\infty}_{k}$ and
\begin{equation*}
\| \F    \|_{L^\infty_{k}}(t) \leq \max\left\{  \| \F_0    \|_{L^\infty_{k}}\, ,  \sum_{i, j=1}^P	 \frac{ 2 \, K}{  \tilde{\eps} }  \,     \|  	f_i    \|_{L^1_{i, k}} \right\}\,,
\end{equation*}
which concludes the proof.
\end{proof}

\begin{remark}
Note that for the case $\tilde{b}_{ij}^{ub} = 1$ in \eqref{p-p ass B}, the propagation result of  $L^\infty$ norms holds for particular gases satisfying $\g/2<\alpha_i$, in the spirit of Remark \ref{Remark const S op}.
\end{remark}

\section*{Acknowledgments}
M.P-\v C. was supported by the Science Fund of the Republic of Serbia, \emph{PROMIS, \#6066089, MaKiPol} and the Ministry of Education, Science and Technological Development of the Republic of Serbia  \emph{\#451-03-68/2022-14/200125}, as well as  by holding an   \emph{Alexander von Humboldt Foundation Fellowship.} 
	
\bibliography{polyatomic-Lp}

\end{document}